\documentclass[aps,pra,reprint,superscriptaddress,10pt,nofootinbib]{revtex4-2}
\usepackage[utf8]{inputenc}
\usepackage[english]{babel}
\usepackage{physics}
\usepackage{times}
\usepackage{graphicx,epsfig}
\usepackage{color}
\usepackage{placeins}
\usepackage[usenames,dvipsnames]{xcolor}
\usepackage{amsmath,bbm,amssymb, amsthm}
\usepackage{dsfont} 
\usepackage{stmaryrd}
\definecolor{linkcolor}{rgb}{0,0,0.6}		
\definecolor{bleu}{HTML}{1732a6}
\usepackage[colorlinks=true,pdfstartview=FitV,linkcolor=linkcolor,citecolor=linkcolor,urlcolor=linkcolor,hyperindex=true,hyperfigures=false]{hyperref}
 
\usepackage{numprint}
\usepackage[normalem]{ulem}

\usepackage{thmtools}
\usepackage{thm-restate}
\usepackage{comment}
\usepackage{hyperref}

\renewcommand{\ketbra}[2]{{|#1 \rangle \langle #2|}}

\newcommand{\s}{\mathbf{s}}

\usepackage{xr}
\makeatletter

\newcommand*{\addFileDependency}[1]{
\typeout{(#1)}
\@addtofilelist{#1}

\IfFileExists{#1}{}{\typeout{No file #1.}}
}\makeatother

\newcommand*{\myexternaldocument}[1]{%
\externaldocument{#1}%
\addFileDependency{#1.tex}%
\addFileDependency{#1.aux}%
}

\myexternaldocument{./Supplemental-Material}

\newtheorem{theorem}{Theorem}

\newtheorem{definition}{Definition}

\begin{document}

\title{Scalable noisy quantum circuits for biased-noise qubits}

\author{Marco Fellous-Asiani}
\email{fellous.asiani.marco@gmail.com}
\affiliation{Centre for Quantum Optical Technologies, Centre of New Technologies, University of Warsaw, Banacha 2c, 02-097 Warsaw, Poland}
\author{Moein Naseri}
\affiliation{Institute of Fundamental Technological Research, Polish Academy of Sciences, \\ Pawi\'nskiego 5B, 02-106 Warsaw, Poland}
\affiliation{Centre for Quantum Optical Technologies, Centre of New Technologies,
University of Warsaw, Banacha 2c, 02-097 Warsaw, Poland}
\author{Chandan Datta}
\affiliation{Institute for Theoretical Physics III, Heinrich Heine University D\"{u}sseldorf, Universit\"{a}tsstra{\ss}e 1, D-40225 D\"{u}sseldorf, Germany}
\affiliation{Department of Physics, Indian Institute of Technology Jodhpur, Jodhpur 342030, India}
\author{Alexander Streltsov}
\affiliation{Institute of Fundamental Technological Research, Polish Academy of Sciences, \\ Pawi\'nskiego 5B, 02-106 Warsaw, Poland}
\affiliation{Centre for Quantum Optical Technologies, Centre of New Technologies,
University of Warsaw, Banacha 2c, 02-097 Warsaw, Poland}
\author{Michał Oszmaniec}
\affiliation{Center for Theoretical Physics, Polish Academy of Sciences, Al. Lotników 32/46, 02-668 Warszawa, Poland}

\begin{abstract}

In this work, we consider biased-noise qubits affected only by  
bit-flip errors, which is motivated by existing systems of stabilized cat qubits. This property
allows us to design a class of noisy Hadamard-tests involving entangling and certain non-Clifford gates, which can be conducted reliably with only a polynomial
overhead in algorithm repetitions. On the flip side we also found classical algorithms able to efficiently simulate both the noisy and noiseless versions of our specific variants of Hadamard test. We propose to use these algorithms as a benchmark of the biasness of the noise at the scale of large circuits. The bias being checked on a full computational task, it makes our benchmark sensitive to crosstalk or time-correlated errors, which are usually invisible from individual gate tomography. For realistic noise models, phase-flip will not be negligible, but in the Pauli-Twirling approximation, we show that our benchmark could check the correctness of circuits containing up to $10^6$ gates, several orders of magnitudes larger than circuits not exploiting a noise-bias. Our benchmark is applicable for an arbitrary noise-bias, beyond Pauli models.
\end{abstract}

\maketitle

\section{Introduction}

Quantum computers bring the hope of solving useful problems for society that would be out of reach from classical supercomputers. One can think of problems in optimisation \cite{Farhi2014Nov,Amaro2022Feb}, cryptography \cite{Shor2006Jul,Haner2020Apr}, finance \cite{chakrabarti2021threshold,rebentrost2018quantum}, quantum chemistry or material sciences \cite{Bauer2020Nov,Cao2019Oct,Ma2020Jul,Cao2018Nov,Zinner2021Jul}. The main threats toward the realization of useful quantum computers are noise and decoherence \cite{Preskill2018Aug} which cause errors and degrade the quality of the computation. In the long term this problem will likely be addressed by
quantum error correction and fault-tolerant quantum computing \cite{Campbell2017Sep, Gottesman2009Apr,Grassl2009,Terhal2015Apr,Aliferis2005Apr,Campbell2017Sep,Fowler2012Aug}. Yet, the very high fidelity and considerable overhead this approach requires makes it very challenging to implement.

In this context, the existence of quantum algorithms able to scale up to large size, with a hardware of low fidelity would be highly desirable. Unfortunately, various studies showed that polynomial-time classical algorithms can efficiently simulate the algorithm outputs of specific circuits in the noisy case, such as random circuits \cite{Aharonov2022Nov,Gao2018Oct} or general algorithms under some assumptions on the noise structure \cite{Aharonov96Noise,DePalma2022Apr,StilckFranca2021Nov} (see also \cite{FeffermanBS2023} and \cite{Brod2020classicalsimulation,GarciaPatron2019simulatingboson} for analogous results in the optical setting). All these studies indicate that without doing error-correction, for realistic noise models, it is not possible to preserve reliable algorithm outputs in a classically intractable regime (see however recent work \cite{Cotler2022}, which showed that in certain oracular scenarios noisy quantum computers can offer an advantage over classical computers). One major difficulty to face is that, for most noise models, the fidelity of the output state drops exponentially with the number of gates involved in the computer \cite{Zhou2020Nov} suggesting that a reliable estimation of any expectation value would require to run exponentially many times the algorithm, ruining any hope for an exponential speedup. Error mitigation techniques \cite{Temme2017Nov,Bonet-Monroig2018Dec,Maciejewski2020mitigationofreadout,Strikis2021Nov,Cai2022Oct,Berg2022Jan,Qin2021Dec,Koczor2021} have been proposed, with the hope to solve this issue. However various no-go results show that, for several noise models, error mitigation techniques are not scalable \cite{Cai2022Oct,Quek2022Oct}: the number of samples they require can grow exponentially with the algorithm's depth or the number of qubits in the algorithm \cite{Takagi2022Aug,Berg2022Jan,Takagi2022Sep}. Other approaches can be potentially more scalable, but they assume specific noise models \cite{Tan2023Jan}, require knowledge of entanglement spectrum of quantum states \cite{EntForging2022} or have potentially high algorithmic complexity \cite{Singal2022}.  

In our work, motivated by the limitations of error-correction and error mitigation, we propose instances of the Hadamard test \cite{Smith2012,Bharti2022Feb} that can be robustly implemented in systems of so-called biased-noise qubits \cite{Puri2020Aug,Guillaud2021Apr,Chamberland2022Feb}, affected only by local stochastic bit flip (or phase-flip) errors. In technical terms, we show that, for suitably designed circuits tailored to a Pauli biased noise  the outcome of the \emph{ideal} version of the Hadamard test can be estimated reliably by execution of its \emph{noisy} versions with only a polynomial overhead in the number of repetitions of the algorithm. The model of biased-noise qubits is motivated by existing hardware realising stabilized cat qubits \cite{Guillaud2021Apr,Lescanne2020May,Chamberland2022Feb}. The key ingredient of our approach is that only specific gates are allowed, in order to preserve the noise bias along the computation \cite{Puri2020Aug,Xu2022Feb}. This allows us to design circuits in which the measurement will be isolated from most of the errors occurring (see Fig. \ref{fig:hadamard_test}). Importantly, the circuits themselves can be quite complicated: the allowed set of gates can generate large amounts of entanglement and contain certain non-Clifford gates, thus avoiding natural classical simulation techniques \cite{Vidal2003,Aaronson2004Nov,Jozsa2003Aug}. Nonetheless, the restricted nature of these circuits allowed us to find an efficient classical algorithm that simulates realisations of our family of Hadamard tests, also in the presence of arbitrary local biased noise. We propose to use this algorithm as a simple benchmark of the biasness of the noise at the scale of large and complicated quantum circuits. The interest of our benchmark is that it is scalable and can detect some collective effects of the noise that cannot, by definition, be observed at the level of individual gates. For instance, it can detect some crosstalk and correlated errors. The overall principle of the benchmark, explained further in section \ref{sec:benchmark}, is to compare the experiment to the classical simulation which assumes the noise model of each gate used inside the complete algorithm is identical to the one deduced from individual gate tomography. Hence, if the simulation and experiment differ, it would imply that collective effects are degrading the quality of the hardware when a complete algorithm is implemented, indicating a potential threat to the scalability of the hardware. For pedagogy, in the main text, we mainly focus on the case of Pauli noise, but our benchmark is applicable for the most general model of local biased model: we provide its result in section \ref{sec:benchmark} and leave the proofs in the supplemental material. 

\section{Methods}

 \subsection{Notations and terminology}
 \label{sec:notations}
 Let $(\sigma_0,\sigma_1,\sigma_2,\sigma_3) \equiv (I,X,Y,Z)$ denote single qubit Pauli matrices. Let $H$ denote the Hadamard gate. We call $\mathbb{P}_n^X$ the set of $X$-Pauli operators acting on $n$-qubits, $\mathbb{P}_n^X \equiv \{ \bigotimes_{k=1}^n \sigma_{i_k}, ~ | \forall k, i_k \in \{0,1\} \}$.
 We say that $f_n \in \text{poly}(n)$ if there exists two reals $C,a>0$ such that $\lim_{n \to \infty }f_n/(C n^a)=1$. Moreover, when $\text{poly}(n)$ appears in an equation, it means that the equation remains true by replacing $\text{poly}(n)$ by any function $f_n \in \text{poly}(n)$. Lastly, we say that $f_n=O(g_n)$ if there exists $C>0$ such that $\lim_{n \to \infty }|f_n/g_n| \leq C$. For any unitary $U$, we define its coherently controlled operation in the $X$-basis as $c_X U \equiv \ketbra{+}{+} \otimes I + \ketbra{-}{-} \otimes U$. Let $G$ be a single-qubit unitary. $G_i$ indicates that $G$ is applied on the $i$'th qubit in the tensor product (and $I$ is applied elsewhere). Calling $\mathcal{G}$ the unitary map implementing a unitary quantum gate $G$, and $\mathcal{E}_{\mathcal{G}}$ the CPTP (Completely Positive Trace Preserving) operation describing the noisy implementation of the gate, we define the "noise map of $G$" (or the noise map associated to $\mathcal{G}$) the CPTP $\mathcal{N}_{\mathcal{G}}$ defined as $\mathcal{N}_{\mathcal{G}} \equiv \mathcal{E}_{\mathcal{G}} \circ \mathcal{G}^{\dagger}$. The noise map of a state preparation is the CPTP map that is applied after a noiseless state preparation, so that the overall process (noiseless preparation followed by CPTP map) describes the noisy state preparation. The noise map of a measurement is the CPTP map that must be applied before the noiseless measurement so that the overall process (CPTP map followed by noiseless measurement) describes the  noisy measurement (noise maps for state preparation and noisy measurements are not uniquely defined).

\label{sec:hadamard_test}
\begin{figure}[h!]
    \centering
    \includegraphics[width=0.9\columnwidth]{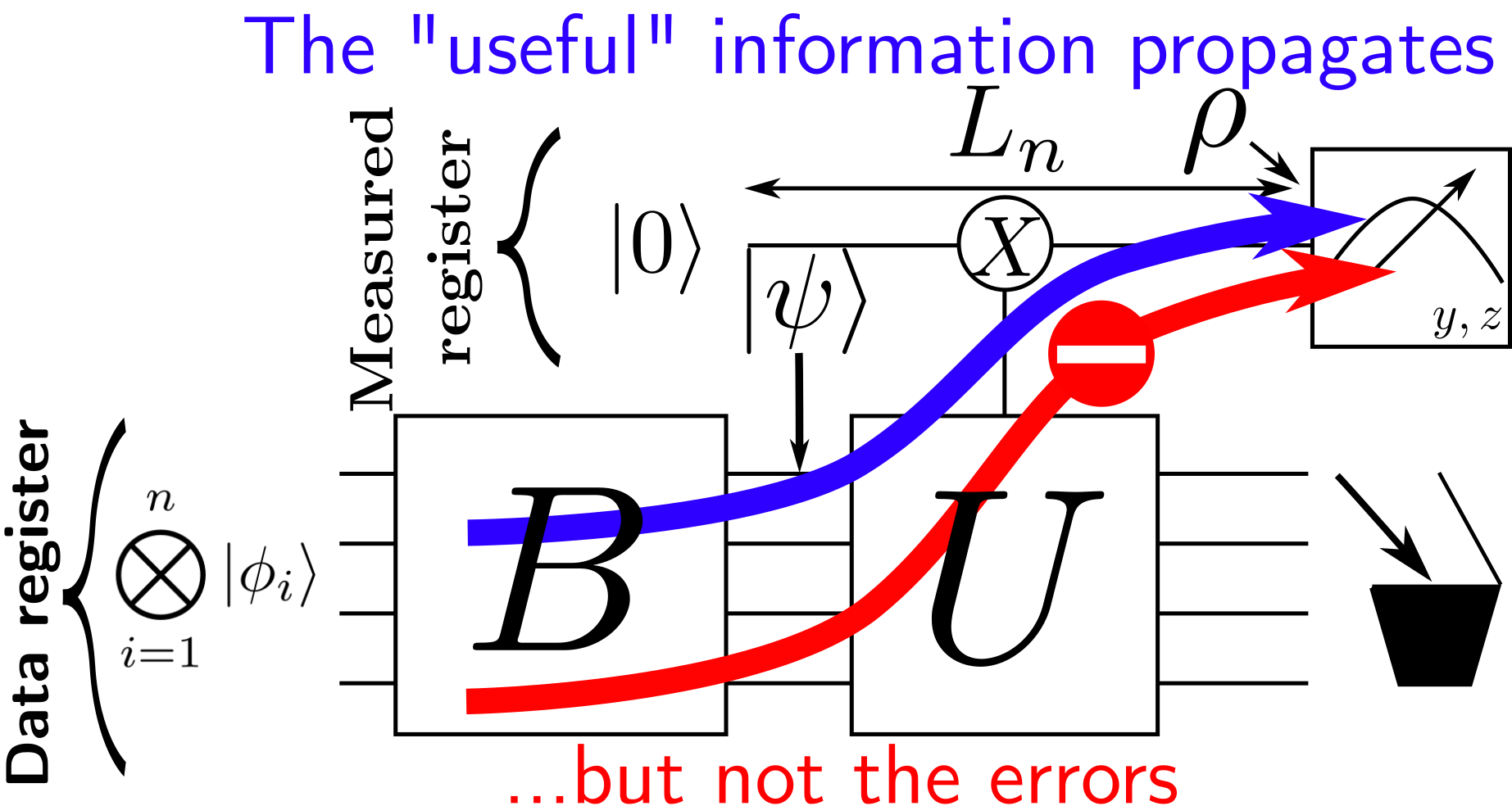}
    \caption{The Hadamard test represented on this figure allows to estimate $\bra{\psi} U \ket{\psi}$ where $\ket{\psi} \equiv B \bigotimes_{i=1}^n \ket{\phi_i}$, ($\ket{\phi_i}$ being a single-qubit state, $B$ a unitary). This estimation is being done by initializing a measured register in $\ket{0}$, and by implementing the unitary $c_X U$. Finally, measuring the measured register in Pauli $Y$ and $Z$ bases gives access to the imaginary and real part of $\bra{\psi} U \ket{\psi}$. We call $L_n$ the number of gates applied on the measured register (including potential noisy identity gates), once $c_X U$ has been decomposed on an experimentally feasible gateset. In this paper, we show that for a noise model only composed of bit-flips, and under some restrictions on the $n$-qubit unitaries $U$ and $B$, the measurement will only be sensitive to bit-flips produced on the measured register, and entirely insensitive to the ones produced on the data register. Nonetheless, some "useful information" contained in the entangled state $\ket{\psi}$ will propagate toward the measurement. The intuitive principle we use is to encode the "useful" information in a "Pauli $Z$ (or $Y$) channel" while making sure that any error is inside a "Pauli $X$ channel". If any Pauli $X$ error (i.e. bit-flip) from the data register cannot propagate toward the measurement while the Pauli $Z$ can, the information will reach the measurements but not the errors. Overall, if $L_n=O(\log(n))$, the noise will only introduce a polynomial overhead in the number of algorithm repetitions to guarantee a reliable outcome:  the algorithm will be scalable despite the noise.
    Note that while $L_n=O(\log(n))$, the algorithm can nonetheless have a polynomial depth.}
    \label{fig:hadamard_test}
\end{figure}

\subsection{The Hadamard test} The Hadamard test is the task over which all our examples are built. It allows to estimate the expectation value $\bra{\psi} U \ket{\psi}$ of a unitary $U$ on some prepared $n$-qubit state $\ket{\psi}=B \bigotimes_{i=1}^n \ket{\phi_i}$ ($\ket{\phi_i}$ being a single-qubit state, $B$ a unitary operation), leading to several applications \cite{Aharonov2009Nov,Gokhale2021Oct,Bharti2022Feb}. A way to implement it is represented in Fig. \ref{fig:hadamard_test}. It consists of a "measured" register initialized in $\ket{0}$ combined with a "data" register, where the state $\ket{\psi}$ has been prepared.
The reduced state of the measured register right before measurement takes the following form
\begin{equation}
    \rho=\frac{1}{2}\left(I+\alpha_n \left( y Y+z Z \right) \right) \, 
    \label{eq:rho_xy_main}
\end{equation}
where $y=-\Im(\bra{\psi} U \ket{\psi}),z=\Re(\bra{\psi} U \ket{\psi})$ and $\alpha_n=1$ for now on. Hence, measuring the first register in the $Y$ (resp $Z$) basis allows to estimate the imaginary (resp real) part of $\bra{\psi} U \ket{\psi}$. Using Hoeffding's inequality \cite{Cai2022Oct} we get that $N=2 \log(2/\delta)/\epsilon^2$ experimental repetitions are sufficient to estimate $y$ or $z$ up to $\epsilon$-precision with a probability $1-\delta$. 
It is well-known that estimating $\bra{\psi} U \ket{\psi}$, for a general polynomial circuit $U$, to additive precision is a BQP complete problem \cite{Aharonov2009Nov} and therefore in general we do not expect an efficient classical algorithm that would be realising this task.

So far, we discussed what happens when the measured qubit is noiseless.  Let us now assume 
that a bit-flip channel right before the measurement occurs in such a way that $0<\alpha_n<1$ in Eq.~\eqref{eq:rho_xy_main}. Then, repeating the algorithm $N_n=2 \log(2/\delta)/(\alpha_n \epsilon)^2$ times would be sufficient to estimate $y$ and $z$ to $\epsilon$ precision with high probability (the scaling in $1/(\alpha_n \epsilon)^2$ is also optimal\footnote{Estimating $y$ (or $z$) up to $\epsilon$ accuracy with a probability greater than $1-\delta$ requires at least $C_{\delta}/(\alpha_n \epsilon)^2$ samples, for some constant $C_{\delta}$. Otherwise it would imply that two Bernoulli distributions of mean $1/2+\alpha_n y$ and $1/2-\alpha_n y$ could be distinguished with a probability higher than $1-\delta$ with fewer than 
$C_{\delta}/(\alpha_n \epsilon)^2$ samples, which is impossible \cite{lectureNotesDistinguishingDistributions}.}). If $\alpha_n$ decreases exponentially with $n$, the total number of algorithm calls will necessarily grow exponentially with $n$ and the algorithm wouldn't be scalable.  However, for $\alpha_n=1/\text{poly}(n)$, only $\text{poly}(n)$ overhead in experiment repetitions would be sufficient to reliably estimate $\bra{\psi} U \ket{\psi}$. In this work, we will show that for biased-noise qubits it is possible to design a class of non-trivial Hadamard tests for which $\alpha_n=1/\text{poly}(n)$ in the presence of non-vanishing local biased noise.

\subsection{Noise model} In general, both $B$, and $c_X U$ have to be decomposed on a gateset implementable at the experimental level. 
Let $\mathcal{G}$ be a unitary channel describing a gate belonging to the accessible gateset, and $\mathcal{E}_\mathcal{G}$ be its noisy implementation in the laboratory. We will assume a local biased noise model: $\mathcal{E}_\mathcal{G} = \mathcal{N}_\mathcal{G} \circ \mathcal{G}$, where the "noise map" $\mathcal{N}_\mathcal{G}$ will only introduce (possibly correlated) bit-flip errors on the qubits on which $\mathcal{G}$ acts non-trivially (i.e. $\mathrm{supp}(\mathcal{G})$),
\begin{equation}
    \mathcal{N}_\mathcal{G}(\rho) =\sum_{\alpha\subset \mathrm{supp}(\mathcal{G})} p^{\mathcal{G}}_\alpha X_\alpha \rho X_\alpha \ ,
    \label{eq:general_noise_model}
\end{equation}
where $X_\alpha =\prod_{i\in\alpha} X_i$ and $\lbrace p^{\mathcal{G}}_\alpha\rbrace$ is a probability distribution supported on subsets of $\mathrm{supp}(\mathcal{G})$. For instance, the noise model of a two-qubit gate will have 
Kraus operators proportional to $\sigma \otimes \sigma'$ with $(\sigma,\sigma') \in \{I,X\}$. Furthermore, a noisy measurement is modelled by a perfect measurement followed by a probability $p_{\text{meas}}$ to flip the outcome. Lastly, we assume that  single-qubit noisy  state preparation consists of a perfect state preparation followed by the application of a Pauli $X$-error with probability $p_{\text{prep}}$. Our noise model is based on an idealization of cat qubits that are able to exponentially suppress other noise channels than bit-flip, at the cost of a linear increase in bit-flip rate \cite{Guillaud2021Apr,Lescanne2020May,Chamberland2022Feb} (or the other way around). It is an idealization as (i) the Kraus operators are a linear combination of Pauli $X$ and $I$ operators only: this is what we call a perfect bias, (ii) we also neglect coherent errors meaning that our noise model is a Pauli noise. The assumption (ii) is mainly used for pedagogical purposes for the main text. Indeed, our main result (the benchmarking protocol) is applicable for biased qubits which are designed in such a manner that their noise model is well approximated by the assumption (i) only (see Theorem \ref{thm:benchmarking} and definition \ref{def:perfect_bias_with_coherent_errors}). We now provide the definition of "an error".  
\begin{definition}[Error]
\label{def:Error}
Let $\ket{\Psi}$ be the state the qubits should be in at some timestep of the algorithm if all the gates were perfect. Because we consider a probabilistic noise model,
the actual $n$-qubit quantum state will take the form 
$\rho = \sum_i p_i E_i \ket{\Psi}\bra{\Psi} E_i^{\dagger}$, where $E_i$ is a unitary operator and $p_i$ some probability ($p_i \geq 0$, $\sum_i p_i=1$). The operator $E_i$ is what we call the error that affected $\ket{\Psi}$.
\end{definition}

\section{Results}
\subsection{Noise-resilient Hadamard test}
The core idea behind our work is to exploit the fact that only bit-flips are produced, in order to design circuits guaranteeing that most of these errors will never reach the measurement in the Hadamard test. In order to explain our results we need to introduce first a number of auxiliary technical definitions.

\begin{definition}[$X$-type unitary operators and errors]
\label{def:X_type_unitary}
We call $X$-type unitary operators the set of unitaries that can be written as a linear combination of Pauli $X$ matrices. For $n$-qubits, we formally define it as:
\begin{align}
    \mathbb{U}_n^X \equiv \{ U=\sum_i c_i P_i, ~ | P_i \in \mathbb{P}_n^X, c_i \in \mathbb{C}, U^{\dagger}=U^{-1} \},
\end{align}
Alternatively, $\mathbb{U}_n^X$ can be understood in terms of unitaries that are diagonal in the product basis $\ket{\s}=\ket{s_1}\ket{s_2}\ldots \ket{s_n}$, where $s_i=\pm$ and $\ket{\pm}$ are eigenstates of Pauli $X$ matrix. We will call "$X$-error", an error belonging to $\mathbb{U}_n^X$ (if the error is additionally Pauli, we will call it "Pauli $X$ error", or simply "bit-flip").
\end{definition}
What we need to do is to guarantee that any error occurring at any step of the computation is an $X$-error. It is possible with the use of "bias-preserving" gates: such gates map any initial $X$-error to another $X$-error. It motivates the following definition and property (all our results are derived in the Supplemental material). 
\begin{definition}[Bias-preserving gates]
\label{def:bias_preserving_gate}
Let $G$ be an $n$-qubit unitary operator. We say that $G$ preserves the $X$-errors (or $X$-bias), if it satisfies the following property
\begin{align}
    \forall P \in \mathbb{P}_n^X, \exists A \in \mathbb{U}_n^X \text{ such that } G P = A G
\end{align}
We denote $\mathbb{B}_n$ the set of such gates. 
\end{definition}
\begin{restatable}[Preservation of the bias]{property}{PresBias}
\label{prop:bias_preservation}
If a quantum circuit is only composed of gates in $\mathbb{B}_n$, each subject to local biased noise model from Eq. \eqref{eq:general_noise_model} (and the paragraph that follows for measurement and preparation), then any error affecting the state of the computation is an $X$-error.
\end{restatable}
As examples of bias-preserving gates, there are all the unitaries in $\mathbb{U}_n^X$, the cNOT, and what we call $\text{Toffoli'} \equiv H_1 H_2 H_3 \times \text{Toffoli} \times (H_1 H_2 H_3)^{\dagger}$\footnote{We assume Toffoli' can be implemented "natively" without having to actually apply the Hadamards (otherwise, because Hadamard would also be noisy, the full sequence wouldn't be bias-preserving)}. An example of a gate that does not preserve the bias is the Hadamard gate. The existence of gate preserving the bias also \textit{during} the gate implementation is not straightforward, but cat-qubits are able to overcome this complication, at least for cNOT, Toffoli' and gates in $\mathbb{U}_n^X$: see the section \textcolor{Blue}{I C} in the Supplemental Material. Finally, bias preserving gates have a nice interpretation: they correspond to permutations (up to a phase) in the Pauli $X$-eigenstates basis. 
 \begin{restatable}[Characterization of bias-preserving gates]{property}{CharBiasPres}\label{lem:BiasPreserv}
$V\in \mathbb{B}_n$ if and only if for any $\s\in \lbrace{+,-\rbrace}^n$, there exists a real phase $\varphi_{\s,V}$ such that: $V \ket{\s}=e^{i \varphi_{\s,V}}\ket{\sigma_V(\s)}$ where $\sigma_V$ is some permutation acting on $\lbrace{+,-\rbrace}^n$.
\end{restatable}
We now sketch the sufficient ingredients guaranteeing the existence of noise-resilient Hadamard tests. First, (i) assume that individual gate errors, as well as individual measurement and state-preparation errors, occur with a probability smaller than $p<1/2$. Furthermore, assume (ii) that only $X$-errors occur in the algorithm (it can be satisfied with the assumptions of Property \ref{prop:bias_preservation}), (iii) these errors cannot propagate from the data to the measured register, (iv) the number of interactions of the measured register with the data register satisfies $L_n=O(\log(n))$ (which implies the measured register will only be impacted by $X$-errors introduced at $O(\log(n))$ locations). Subject to conditions (i-iv) the reduced state $\rho$ will satisfy Eq. \eqref{eq:rho_xy_main} with $\alpha_n$ efficiently computable classically, and satisfying $\alpha_n \geq 1/\text{poly}(n)$. As explained earlier, this would guarantee a scalable algorithm to estimate $\bra{\psi} U \ket{\psi}$. In particular, the following Theorem holds. 
\begin{restatable}[Hadamard test resilient to biased noise]{theorem}{HadamResNoise}
\label{app:thm:noise_resilience}
Let:
\begin{align}
& \ket{\psi}=B \bigotimes_{i=1}^{N_B} \ket{\phi_i}, \ U=W \cdot V\ , \notag \\
&\ W \equiv \prod_{i=1}^{N_W} W_i, \ V \equiv \prod_{i=1}^{N_V} V_i\ ,
\label{eq:U_main_result}
\end{align}
where $B$ is a product of local bias preserving gates, gates $V_i$ and $W_i$ are local gates and belong to $\mathbb{U}_n^X$. Additionally, the gates $W_i$ are assumed to be Hermitian. We assume the circuit is implemented as indicated on Figure \ref{fig:parallelisation_register}. There, $W$ is implemented thanks to the "parallelisation register", while $V$ is implemented by making the measured and data register directly interact.

Furthermore, we assume the local bias noise model introduced in Eq. \eqref{eq:general_noise_model}, and that state preparation, measurements, and each non-trivial gate applied on the measurement register have a probability at most $p<1/2$ to introduce a bit-flip on the measured register.

Under these conditions, there exists a quantum circuit realising a Hadamard test such that, in the presence of noise, the reduced state $\rho$ satisfies Eq.~\eqref{eq:rho_xy_main} with $\alpha_n \geq (1-2 p)^{O(N_V)}$.

Additionally, $\alpha_n$ is efficiently computable classically. Hence, if $N_V=O(\log(n))$, it is possible to implement the Hadamard test in such a way that running the algorithm $\text{poly}(n)$ times is sufficient to estimate the real and imaginary parts of $\bra{\psi} U \ket{\psi}$ to $\epsilon$ precision with high probability.
\end{restatable}
The Hadamard test is implemented with the circuit shown in Figure \ref{fig:parallelisation_register}. It uses a parallelization register. It is this parallelization register that allows to implement a unitary acting on all the qubits of the data register (the unitary $W$ mentionned in the theorem), while keeping a logarithmic number of interactions with the measured register (this is required to preserve a noise-resilience). While Theorem \ref{app:thm:noise_resilience} assumes the trivial identity gate to be noiseless, the results can be easily extended if they are noisy (in this case, $W$ should have a depth in $O(\log(n))$, as discussed in the section \textcolor{Blue}{III C} of the Supplemental Material). Our approach is also scalable in the presence of noisy measurements (measuring the data register in the $X$ basis to infer $\bra{\psi} U \ket{\psi}$ would not be scalable in general in this case: see \cite{Obrien20022} and the section \textcolor{Blue}{IV B} of the Supplemental Material).

\subsection{Efficient simulation of noise-resilient Hadamard test}
We just saw that restricted forms of Hadamard tests are resilient to bit-flip errors occurring throughout the circuit. The following result proves that the task realised by such a restricted Hadamard test can be efficiently simulated on a classical computer with polynomial effort.
\begin{restatable}[Efficient classical simulation of restricted Hadamard test]{theorem}{EfficientSim}
\label{thm:simulation}
Let $B\in\mathbb{B}_n, U\in\mathbb{U}^X_n$ be n qubit unitaries specified by $R_B$ and $R_U$ local qubit gates (belonging to respective classes $\mathbb{B}_n$ and $\mathbb{U}^X_n$). Let $\ket{\psi_0}=\ket{\phi_1}\ket{\phi_2}\cdots\ket{\phi_n}$ be an initial product state. Then, there exists a randomized classical algorithm $\mathcal{C}$, taking as input classical specifications of circuits defining $B$, $U$, and the initial state $\ket{\psi_0}$, that efficiently and with a high probability computes an additive approximation to $\bra{\psi_0} B^\dagger U B \ket{\psi_0}$. Specifically, we have 
\begin{equation}
    \Pr\left( |\bra{\psi_0} B^\dagger U B \ket{\psi_0} -\mathcal{C}| \leq \epsilon \right)\geq 1-\delta \ ,
\end{equation}
while the running time is $T=O\left(\frac{R_B + R_U +n}{\epsilon^2} \log(1/\delta)\right)$.
\end{restatable}

The classical simulability of the restricted Hadamard test can be regarded as the limitation of our approach to construct quantum circuits that are robust to biased noise. At the same time, it allows us to introduce an efficient and scalable benchmarking protocol that is tailored at validating the assumption of bias noise on the level of the whole (possibly complicated) circuit. It allows to validate this assumption in a manner that individual gate tomography couldn't, as we now explain. 

\subsection{Benchmarking the biased noise model at the scale of a whole circuit}
\label{sec:benchmark}
\subsubsection{Sketch for Pauli bit-flip noise}
The issue with an individual benchmark of quantum gates is that it is blind to collective effects that can only build up when a full circuit, composed of many gates in sequence and in parallel, is implemented. One example is the presence of correlated errors, or more generally, non-local effects in the noise, that individual gate tomography cannot usually detect. Another example of collective effects is the presence of "scale-dependent noise", i.e. a noise for which the intensity depends on the number of qubits or surrounding gates used in the algorithm \cite{Zhao2022Apr,krinner2020benchmarking,sevilla2020forecasting}. Note that scale-dependent noise can sometimes be implied by correlated noise models \cite{Fellous-Asiani2021Nov}. Scale-dependent noise can be particularly damaging for biased qubits as it could rule out the intrinsic interest of using such qubits in the case the nearly suppressed error rate happens to grow when the circuits are scaled up. All these behaviors for the noise represent a major threat both for the near term, and to reach the large-scale \cite{murali2020software,agarwal2023modelling,zhao2022quantum}. Our benchmarking protocol, that we now sketch, allows to detect some of these effects. The intuitive idea is that we are able to classically predict the algorithm's output under the assumption that the noise deduced from individual gate tomography is still realized for each gate but now inside of a whole algorithm. A mismatch between the classical algorithm and the quantum experiment would necessarily indicate a violation of the assumption behind the noise. 

\begin{figure}[h!]
    \centering
\includegraphics[width=\columnwidth]{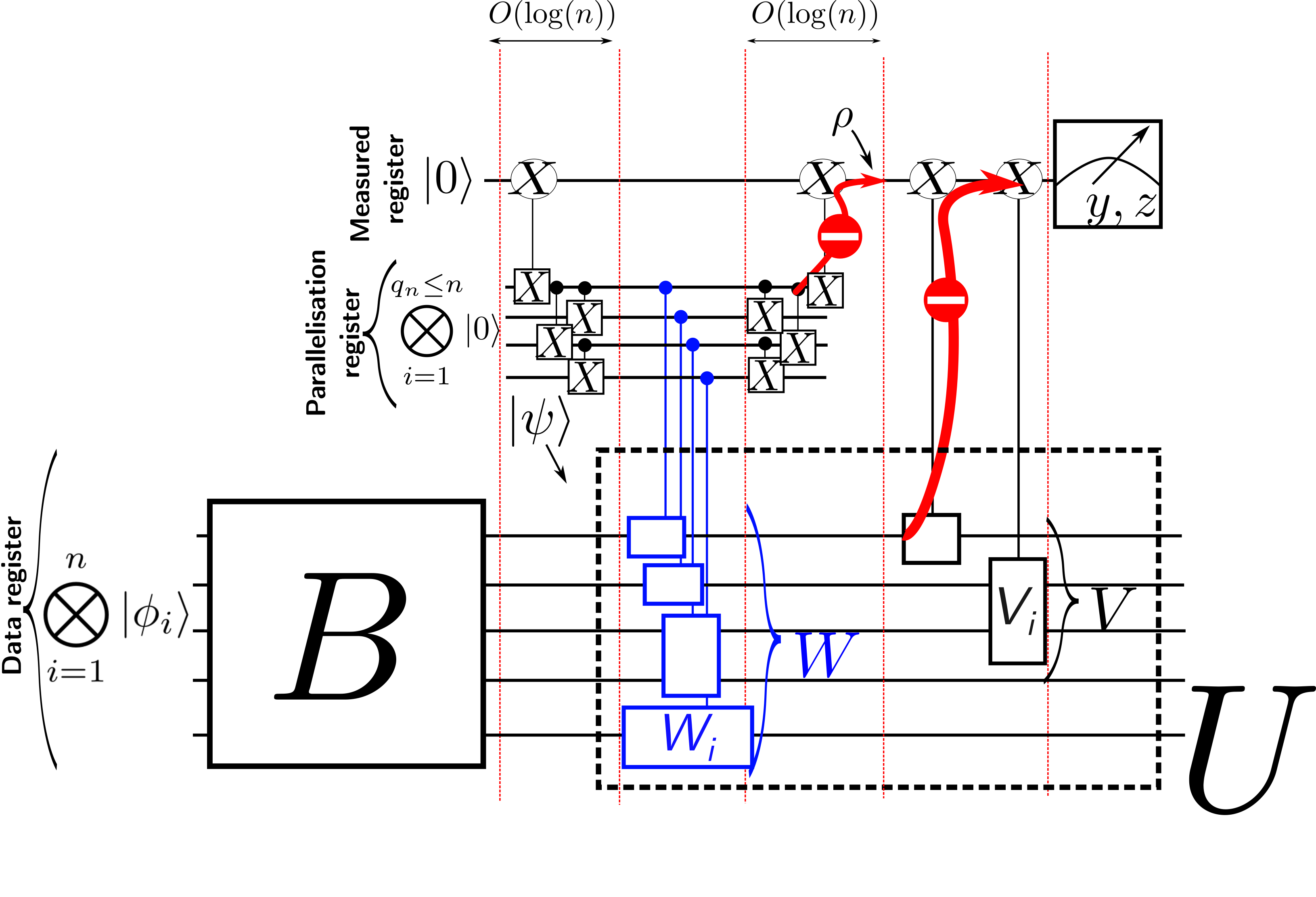}
    \caption{In this figure, we illustrate how $U=W \times V = \bigotimes_{i=1}^{N_W} W_i \times \bigotimes_{i=1}^{N_V} V_i$ mentionned in theorem \ref{app:thm:noise_resilience} can be implemented in a noise-resilient manner. The coherently controlled unitaries with a blue contour implement the controlled Hermitian unitaries $W_i$ which overall implement the controlled $W$. The coherently controlled unitaries with a black contour implement the controlled $V_i$, which overall implement the controlled $V$. The central element in this construction is the parallelisation register \cite{moore2001parallel}, which allows to implement the unitary $W$ in such a way that it acts upon all the qubits in the data register while preserving the noise-resilience. Hence, it allows to implement a Hadamard test with $U$ acting on all the qubits of the data register. Such unitaries are useful for our benchmarking protocol: see the third paragraph in the discussion section. The key point behind the parallelisation register is that, while it introduces additional components that can introduce bit-flip errors, these bit-flips cannot, by construction, propagate to the measured register (in practice, they commute with the last $c_X X$ gate drawn). It is an example where trading space (using more qubits) to gain time (guaranteeing that the measured register interacts with $O(\log(n))$ gates and not $\text{poly}(n)$ gates) is worth doing. While the parallelisation register does not propagate bit-flips toward the measured register, it will propagate phase-flip errors. We make use of this to easily detect an excessive production of phase-flip errors in the third paragraph of the discussion section. Note that, while the evaluation of $\bra{\psi}U\ket{\psi}$ requires the whole circuit of figure \ref{fig:parallelisation_register} to be noise-resilient for the bit-flip noise described around \eqref{eq:general_noise_model} (see \cite{footnote-eq_circuit})), as discussed at the end of the third paragraph of the discussion section, it could be that simpler circuits than the ones of Figure \ref{fig:parallelisation_register} could be used for the sole purpose of benchmarking, depending on which violation in the assumed noise model we want the benchmark to be sensitive to. However, to keep our explanations simple, we have chosen to exclusively focus on the circuit of Figure \ref{fig:parallelisation_register} all along our paper.}
\label{fig:parallelisation_register}
\end{figure}

For pedagogy, we first give a sketch of how the benchmarking would work for the Pauli bit-flip noise model considered so far. Then, we will give our main result which is a benchmarking protocol working for the most general case of biased qubits. Namely, it is a benchmarking working for qubits which noise model should correspond to a perfect bias which is not necessarily Pauli, i.e., a noise model satisfying the definition \ref{def:perfect_bias_with_coherent_errors}. Note that, for simplicity of the explanations, in all that follows we propose to implement our benchmark in the exact circuits we analyzed so far (i.e. the ones of Figure \ref{fig:parallelisation_register}). However, as mentionned at the end of the third paragraph of the discussion section, depending on which violation in the noise we aim to be sensitive to, simpler circuits could potentially be considered.

For a Pauli bit-flip noise, we first need to perform tomography on the gates that need to be applied on the measured register, in order to find the probability that each such gate introduces a bit-flip on the measured register. Other noise parameters $p_{\text{meas}}$  and $p_{\text{prep}}$ would also need to be estimated. Then, a Hadamard test satisfying the constraints of Theorem \ref{app:thm:noise_resilience} is chosen. Subsequently, one needs to classically compute $\alpha_n$, and estimate $y,z$ up to $\epsilon$-precision, with high probability  (using  Theorem \ref{thm:simulation}). The estimates of the \emph{noiseless} $y,z$ are compared to values of $Tr(\rho Y)/\alpha_n$, $Tr(\rho Z)/\alpha_n$ estimated experimentally by running the quantum circuit polynomially many times. The experimental and theoretical predictions are then compared. If they don't match (up to error resulting from a finite number of experiments), it would necessarily indicate a violation of the assumptions behind the assumed noise model (bit-flip) for the circuit participating in the Hadamard test.
\subsubsection{General case}
In general, the noise model describing biased qubits can have its Kraus operators describing the noise map of each gate, state preparation and measurements that do not exactly correspond to a Pauli bit-flip channel. We mean that the Kraus operators could be a linear combination of Pauli bit-flips (i.e. they would follow the definition \ref{def:perfect_bias_with_coherent_errors}: the bit-flip channel treated so far was a particular case of the expected noise model for such hardware). Additionally, there can have imperfection in the bias. It means that the Kraus operators describing the noise maps could have a non-zero Hilbert-Schmidt inner product with Pauli operators not belonging to $\mathbb{P}_n^X$. For instance, the Kraus operators could have a non-zero overlap with a multi-qubit Pauli $P$ containing at least one Pauli $Z$, or one Pauli $Y$ in the expression of its tensor product (such as $P=X \otimes Z$ for instance). If it occurs, we will say that the noise model also produces phase-flip errors. Our benchmarking protocol can detect violations of the assumed noise model in this more general case. The protocol is based on the fact that we can simulate the outcome of the Hadamard test in the \textit{noisy} case, under the assumption that the bias is perfect but not necessarily Pauli (i.e. that it satisfies definition \ref{def:perfect_bias_with_coherent_errors}). This classical simulation takes as input the noise model of each individual gate approximated to the one of a perfect bias. A deviation of this classical simulation to the experimental outcomes further than an error budget will imply that the noise model prescribed by individual gate tomography is not occuring experimentally, identifying a possible threat to the hardware scalability, due to collective effects occuring at the scale of the algorithm. This error budget is related to the quality of the approximation of the noise model taken from individual gate tomography by the one of a perfect bias. The formal protocol is written in theorem \ref{thm:benchmarking}. We now state the definition and theorems we need.
\begin{definition}{Perfect bias}

\label{def:perfect_bias_with_coherent_errors}
    Let $\mathcal{G}$ be a quantum channel describing either a noiseless unitary gate belonging to the accessible gateset, or a single-qubit state preparation, and $\mathcal{E}_\mathcal{G}$ be its noisy implementation in the laboratory. We say that the noisy implementation of the gate (or state preparation), $\mathcal{E}_\mathcal{G}$, follows a perfectly biased noise model if $\mathcal{E}_\mathcal{G} = \mathcal{N}_\mathcal{G} \circ \mathcal{G}$, where the noise map $\mathcal{N}_\mathcal{G}$ is a CPTP (Completely Positive Trace Preserving) map that admits the following Kraus decomposition:
\begin{align}
    &\mathcal{N}_\mathcal{G}(\rho) =\sum_{j} K^{\mathcal{G}}_{j} \rho (K^{\mathcal{G}}_{j})^{\dagger} \ , \notag \\
    &K^{\mathcal{G}}_{j}=\sum_{\alpha \subset \mathrm{supp}(\mathcal{G})} c^j_\alpha X_\alpha,
    \label{eq:noise_model_with_coherent_errors}
\end{align}
where $X_\alpha =\prod_{i\in\alpha} X_i$, $\forall j, c^j_\alpha \in \mathbb{C}$ and $\sum_j (K^{\mathcal{G}}_j)^{\dagger} K_j^{\mathcal{G}}=I_{\mathrm{supp}(\mathcal{G})}$, $I_{\mathrm{supp}(\mathcal{G})}$ being the identity operator applied on the qubits where $\mathcal{G}$ acts non-trivially. A quantum map satisfying \eqref{eq:noise_model_with_coherent_errors} will be said to be perfectly biased.

A noisy single-qubit measurement will be modelled as a perfect measurement, preceded by the application of a perfectly biased single-qubit CPTP map on the qubit being measured. This CPTP channel, called the noise map of the measurement, represents the noise bringed by the measurement.
\end{definition}
To make the formulation of the following theorems simple, we will assume that state preparation and measurements are noiseless. In the case all gates are bias-preserving (which is what we consider in all this paper) and if the noise model satisfy definition \ref{def:perfect_bias_with_coherent_errors}, this assumption doesn't remove generality to our results. This is because the noise of the state preparation or measurement can be acknowledged by redefining the noise map of the following or preceding gate. The newly obtained noise map will still satisfy definition \ref{def:perfect_bias_with_coherent_errors}.  See \cite{footnote-redefine-noise} for more details.


\begin{theorem}{Efficient classical simulation of a noisy Hadamard test under perfect bias.}

\label{thm:simulation_extended}
Let $B \in \mathbb{B}_n$, $U=W.V \in \mathbb{U}_n^X$, $N_W, N_V$ satisfying the constraints of Theorem \ref{app:thm:noise_resilience}. Let $\ket{\psi_0}=\ket{\phi_1}\ket{\phi_2}\cdots\ket{\phi_n}$ be the initial product state for the data register. If $N_V=O(\log(n))$, if the noise model of each gate is perfectly biased according to definition \ref{def:perfect_bias_with_coherent_errors}, if the total number of gates in the algorithm is in $\text{Poly}(n)$, and if state preparation and measurements are noiseless (this is without loss of generality, see the comments before the Theorem \ref{thm:simulation_extended}), there exists a randomized classical algorithm $\mathcal{C}$ taking as input (I) classical specifications of the circuit implementing the Hadamard test for the chosen $(B,U,\ket{\psi_0})$, (II) the quantum channel describing the noise model of each gate used in the computation, (III) the initial state $\ket{\psi_0}$ such that $\mathcal{C}$ efficiently and with a high probability computes an additive approximation to $Tr(P_1 \rho_X)$, where $\rho_X$ is the reduced density matrix of the measured register at the end of the noisy implementation of the algorithm, and $P_1$ is a single-qubit Pauli matrix. 

Specifically, we have:
\begin{align}
    \Pr\left( |Tr(P_1 \rho_X) -\mathcal{C}| \leq \epsilon \right)\geq 1-\delta \ ,
\end{align}
while the running time is $T=O((1/\epsilon^2) \log(1/\delta) \times \text{poly}(n))$.
\end{theorem}


\begin{restatable}[Benchmarking protocol]{theorem}{benchmarking}
\label{thm:benchmarking}
Consider a Hadamard test satisfying the constraints of Theorem \ref{app:thm:noise_resilience}. The circuit is implemented with $N$ unitary gates represented by the unitary quantum channels $\{\mathcal{G}_i\}_{i=1}^N$. We call $\mathcal{N}_{\mathcal{G}_i}$ the noise map associated to $\mathcal{G}_i$ (extracted from individual gate tomography) and we assume state preparation and measurements to be noiseless (this is without loss of generality up to a redefinition of the noise maps of the quantum gates, see comments before Theorem \ref{thm:simulation_extended}).

Let $\mathcal{N}_{X,\mathcal{G}_i}$ be an approximation to $\mathcal{N}_{\mathcal{G}_i}$, such that $\mathcal{N}_{X,\mathcal{G}_i}$ has a noise model satisfying definition \ref{def:perfect_bias_with_coherent_errors}.

Let $\rho$ be the density matrix of the measured register at the end of the algorithm if the noise map of each $\mathcal{G}_i$ was $\mathcal{N}_{\mathcal{G}_i}$. Let $\rho_X$ be the reduced density matrix of the measured register at the end of the algorithm if the noise map of each $\mathcal{G}_i$ was $\mathcal{N}_{X,\mathcal{G}_i}$. Let $\rho_{\text{exp}}$ be the reduced density matrix of the measured register that exactly predicts the experimental outcomes. More precisely, we mean that implementing the measurements used in the Hadamard test on $\rho_{\text{exp}}$ would exactly reproduces the measurement outcomes experimentally observed.

Assume that there exist $\epsilon>0$ such that 
\begin{align}
    \max_{i} ||\mathcal{N}_{X,\mathcal{G}_i}-\mathcal{N}_{\mathcal{G}_i}|| \leq \frac{\epsilon}{\sqrt{2} N}
    \label{eq:NepsilonBenchmarking}
\end{align}
is satisfied. Then, for any single-qubit Pauli $P_1$:
\begin{align}
    |Tr(\rho P_1)-Tr(\rho_X P_1)| \leq \epsilon
    \label{eq:Tr_diff}
\end{align}

\textbf{Principle of the benchmarking:}

The benchmarking protocol works as follow. $Tr(\rho_X P_1)$ corresponds to the outcome of the circuit with a noise model for each gate satisfying definition \ref{def:perfect_bias_with_coherent_errors}. Hence, it can be classically estimated with Theorem \ref{thm:simulation_extended}. $Tr(\rho_{\text{exp}} P_1)$ is accessed experimentally. If $|Tr(\rho_{\text{exp}} P_1)-Tr(\rho_{X} P_1)|>\epsilon$, $\rho_{\text{exp}}$ and $\rho$ necessarily differ, indicating that the noise model predicted by individual gate tomography (the noise maps $\{\mathcal{N}_{\mathcal{G}_i}\}_{i=1}^N$) is not occuring experimentally. It thus indicates the presence of collective effects in the noise, possibly threatening the scalability of the biased-noise qubits.

Noteworthy, $\Delta \equiv |Tr(\rho_{\text{exp}} P_1)-Tr(\rho_X P_1)|-\epsilon$ can quantify how strong the noise violation is at the scale of the whole circuit (if $\Delta > 0$). It is a consequence of the fact $|Tr(\rho_{\text{exp}} P_1)-Tr(\rho P_1)| \geq \Delta$ (the larger $\Delta$, the larger the violation).
\end{restatable}
\section{Discussion}
In supplemental \textcolor{Blue}{V B.3}, we quantify the total number of quantum gates allowed in the circuit for which our benchmark could be practically useful. Based on available data in literature, we show that under the Pauli-Twirling approximation, it could be used for circuits containing $10^6$ gates. As it represents circuits three to four orders of magnitude bigger than in current experiments \cite{pelofske2022quantum}, it is a strong indication of the practical usefulness of our benchmark for the NISQ regime, and beyond, allowing to check the hardware reliability for large circuits. Loosely speaking, our benchmark shall be used in the regime where the suppressed source of errors (phase-flip in our convention) is expected to be negligible. This is because it relies on the classical simulation, theorem \ref{thm:simulation_extended}, which assumes a perfect bias, hence that no phase-flip are produced in the algorithm. This is precisely the regime where having a benchmark is interesting as there are propositions to use biased-qubits in large-scale algorithms where only the dominant source of errors is corrected, because the suppressed errors would have a negligible impact at the level of the algorithm (see supplemental E2 of \cite{gouzien2023performance} and references therein). Hence, our protocol is well-suited to analyze this regime of interest.

Our protocol can first detect some correlated errors or, more generally, non local effects of the noise that are usually invisible from individual gate tomography. This is because our classical simulation algorithm assumes that individual gate tomography provides a fair description of the behavior of the noise, as it takes as inputs the noise maps of the gates extracted from tomography. Thus, if gate tomography does not fairly describe the noise, the results from simulation and experiment would disagree. To be concrete: if the noise model satisfies definition \ref{def:perfect_bias_with_coherent_errors}, every gate applied on the measured register introduces noise that can impact the measurement outcomes. Furthermore, these are the only gates for which the noise can impact the measurement outcomes.  This is because, with a perfect bias, no error produced outside of the measured register can propagate to the measured register (this is a consequence of how our circuits are built). Hence, we can first detect the presence of some correlated errors between the gates applied on the measured register as they would violate the assumption that each gate can be described with a noise map independent from each other: this is a first example of a violation we can detect. We can also detect some violations of the locality assumption of the noise. If there are non-local effects in the noise, the noise maps of the gates applied on the measured register, would, in general, not correctly describe the experimental outcomes. For instance, it could be that some operations done locally in the parallelisation or data registers introduces unexpected errors on the measured one (because of crosstalk for example). Our benchmarking will, in general, allow to detect such effects. These are just some examples to help the reader understand what the benchmark can be useful for. Yet, we think that the most useful application of our protocol is its ability to detect the occurence of phase-flip errors produced at a higher rate than expected (we recall that having error-rates growing with the computer's size is an effect experimentally observed in superconducting qubits \cite{Zhao2022Apr,krinner2020benchmarking,sevilla2020forecasting}). Indeed, such effects would in general lead to a mismatch between the classical simulation and the experiment for algorithms containing less gates than expected (i.e. a total number of gates, $N$, smaller than what \eqref{eq:NepsilonBenchmarking} predicts). To be concrete, an experimentalist should choose the largest $N$ so that the bound \eqref{eq:NepsilonBenchmarking} is saturated. If $\Delta \geq 0$, there is a violation of the assumption behind the noise model, indicating a potential threat to the scalability of the platform.  For instance, for $\epsilon=1/50$, our quantitatives estimates of supplemental \textcolor{Blue}{V B.3} indicate that $N=10^6$ would work.

We can also provide a concrete example of circuit that can be used to efficiently detect the production of phase-flip at a higher rate than expected. It would be the one implementing the controlled unitary $c_X U$, with $U=\otimes_{i=1}^n X_i$. This circuit matches the constraints of Theorem \ref{app:thm:noise_resilience} (including its generalisation to noisy identity gates mentioned in the paragraph that follows Theorem \ref{app:thm:noise_resilience}): we can thus implement it for the benchmarking. The reason why this circuit is useful is that it would in general make the measurement outcome sensitive to the introduction of phase-flip error on any of the qubits from the data register, after the preparation unitary $B$. This is because a phase-flip error occuring on any of the qubits from the data register would first propagate to the parallelisation register (through the blue cNOTs implementing $W=U=\otimes_{i=1}^n X_i$ in Figure \ref{fig:parallelisation_register}), and finally to the measured register. Look at \cite{footnote-propaZ} for a more detailed explanation. Hence, phase-flip errors produced in $B$ will in general modify the measurement outcome probability distribution, allowing to efficiently detect an excessive production of such errors \cite{caveat-protocol}. An experimentalist could then see if the circuit composed of bias-preserving gates of its interest, encoded in the unitary $B$, does indeed produce more phase-flip errors than it should. Detecting such events is crucial for superconducting cat qubits given the fact their whole scalability strategy precisely relies on keeping negligible the phase-flip error rates also when used in large-scale circuits \cite{gouzien2023performance}. Note that if the goal is only to detect the production of phase-flips at a higher rate than expected, and not to see if other non-local effects are occuring (we can detect some of these other non-local effects with our benchmarking, see the previous paragraph), simpler circuits than the ones based on Figure \ref{fig:parallelisation_register} could very likely be used for benchmarking. Assuming noiseless measurements, one could for instance measure the observable $X^{\otimes n}$ on the prepared $\ket{\psi}$ by directly measuring all the data qubits in the Pauli-$X$ basis (i.e., in this case, one would remove the parallelisation and measured register of Figure \ref{fig:parallelisation_register}). In presence of a perfect bias, the errors will commute with the measurements. Hence, the implementation of the circuit would give the same outcome as the classical simulation algorithm Theorem \ref{thm:simulation} (for $U=X^{\otimes n}$, $B \ket{\psi_0} = \ket{\psi}$). For this reason, we believe that a similar benchmarking protocol as Theorem \ref{thm:benchmarking} could be derived for this simpler circuit (a mismatch between the simulation and experiment would indicate a violation). However, we leave a rigorous proof of this guess for a future work (the proof should in particular acknowledge what happens if these measurements are noisy -- something we acknowledged in Theorem \ref{thm:benchmarking}\footnote{See the paragraph preceeding Theorem \ref{thm:simulation_extended} and the noise of definition \ref{def:perfect_bias_with_coherent_errors}. This noise model makes Pauli-$Y$ and Pauli-$Z$ measurements noisy, which are the only ones we use in our circuit of Figure \ref{fig:parallelisation_register}} -- and precisely quantify the violation with similar equations as \eqref{eq:NepsilonBenchmarking} and \eqref{eq:Tr_diff}).

Overall, the important aspect of our benchmark is that it is scalable (even if measurements and state preparation are noisy), and allows to detect violation of the noise model in a manner that would not be visible from individual gate tomography, because some effects of the noise cannot be detected at this level, as we discussed at the beginning of section \ref{sec:benchmark}.

Finally, while our circuits are efficiently simulable, their strong noise-resilience, rigorously proven in presence of a Pauli bit-flip noise (i.e. the noise model described around \eqref{eq:general_noise_model})\footnote{For a non-Pauli noise model exactly satisfying the definition \ref{def:perfect_bias_with_coherent_errors}, it would still be true that the only noise damaging the measurements would be the one produced by the gates applied on the measured register. However, this noise could, in general, modify the measured register density matrix in a manner that doesn't allow to recover the noiseless evaluation of the Hadamard test. This is however not an issue for our benchmarking.} makes natural to wonder if extensions of our work could lead to noise-resilient circuits also showing a computational interest: this work also shows what we analyzed on this question of fundamental interest. For this, we first notice that the set of bias-preserving gates, $\mathbb{B}_n$ can be composed of $c_X X$ gates, which combined to initial states $\ket{0/1}$ can generate arbitrary graph states (in the local $X$ basis). It is known that a typical $n$-qubit stabilizer state exhibit strong multipartite entanglement \cite{smith2006typical}. This, together with the fact that arbitrary graph states are locally equivalent to stabilizer states \cite{Stab2004}, implies that, in general bias-preserving circuits can generate a rich family of highly entangled states (these graphs states are however not computationally useful for the specific task adressed in our paper, see the section \textcolor{Blue}{II B} of the Supplemental Material). We can also implement certain non-Clifford gates, and for our task, $U \in \mathbb{U}_n^X$ can act non-trivially over all the qubits of the data register, there is no restriction on the number of gates in the preparation unitary $B$, and the circuit is scalable despite noisy measurements. The error rates of the gates could also grow with $n$, making our circuits noise-resilient against one of the main threats to the scalability \cite{Fellous-Asiani2021Nov}. (see section \textcolor{Blue}{IV A} of the Supplemental Material).

\begin{acknowledgments}
\textit{Acknowledgments}: This work benefits from the ``Quantum Optical Technologies'' project, carried out within the International Research Agendas programme of the Foundation for Polish Science co-financed by the European Union under the European Regional Development Fund and the ``Quantum Coherence and Entanglement for Quantum Technology'' project, carried out within the First Team programme of the Foundation for Polish Science co-financed by the European Union under the European Regional Development Fund. MO acknowledges financial support from the Foundation for Polish Science via TEAM-NET project (contract no.\ POIR.04.04.00-00-17C1/18-00). CD acknowledges support from the German Federal
Ministry of Education and Research (BMBF) within the funding program ``quantum technologies -- from basic research to market'' in the joint project QSolid (grant
number 13N16163). This project also received funding from the European Union's Horizon Europe Research and Innovation programme under the Marie Sklodowska-Curie Actions \& Support to Experts programme (MSCA Postdoctoral Fellowships) - grant agreement No. 101108284. The authors thank Daniel Brod, Jing Hao Chai, Jérémie Guillaud, Michal Horodecki, Hui Khoon Ng and Adam Zalcman for useful discussions that greatly contributed to this work.
\end{acknowledgments}

\bibliography{main}

\begin{thebibliography}{72}
\bibitem [{foo({\natexlab{b}})}]{footnote-eq_circuit}%
  \BibitemOpen
  \href@noop {} {} ({\natexlab{b}}),\ \bibinfo {note} {{Directly measuring the data qubits in the Pauli $X$-basis to infer $\bra{\psi}U\ket{\psi}$ would, in general, not be noise resilient: see the paragraph after Theorem \ref{app:thm:noise_resilience}.}}\BibitemShut {Stop}%
\bibitem [{cav()}]{caveat-protocol}%
  \BibitemOpen
  \href@noop {} {}\bibinfo {note} {{We say "in general", because depending on the implemented $B$, $Z$ errors (i.e. phase-flip) occurring at specific locations could be left undetected. For instance, if the overlap measured by the Hadamard test, $\bra{\psi} U \ket{\psi}$ satisfies $\bra{\psi} U \ket{\psi}= \bra{\psi}Z_i U Z_i \ket{\psi}$, a $Z$ error produced on the $i$'th data qubit right before the implementation of $U$ would have no impact on the measurement probability distribution.}}\BibitemShut {Stop}%
\end{thebibliography}
\end{document}


\begin{center}
\textbf{\large Supplemental Materials}
\end{center}
\section*{General comments before reading this supplemental material}
In this work, we considered qubits having a noise model satisfying definition \textcolor{Blue}{4} of the main text. In the existing literature, another convention is often taken: there the dominant source of errors is usually phase-flip rather than bit-flips. It means that their noise model satisfy the definition \textcolor{Blue}{4} of the main text, where Pauli $X$ are replaced by Pauli $Z$.

To simplify the writing, in some parts of this supplemental material, we will say that the algorithm (Hadamard test) is scalable. It is a short way to say that one could estimate $\bra{\psi} U \ket{\psi}$ up to $\epsilon$ precision with a probability $1-\delta$, with a total number of operations (repetitions of the algorithm + the quantum gates used in the algorithm) that scales polynomially with $n$, the problem size.

\section{Bias preserving gates and error propagation}
Here, we will characterize the unitaries belonging to $\mathbb{B}_n$. We will also characterize the controlled unitaries that do not propagate bit-flip errors from the target to the control.
\subsection{Arbitrary bias-preserving gates}
We will prove the following Property that appears in the main text.
\label{app:general_unitaries}
\setcounter{property}{1}
\begin{property}[Characterization of bias-preserving gates]\label{lem:BiasPreserv}
$V\in \mathbb{B}_n$ if and only if for any $\s\in \lbrace{+,-\rbrace}^n$, there exists a real phase $\varphi_{\s,V}$ such that: $V \ket{\s}=e^{i \varphi_{\s,V}}\ket{\sigma_V(\s)}$ where $\sigma_V$ is some permutation acting on $\lbrace{+,-\rbrace}^n$.
\end{property}
\begin{proof}
We consider $\s\in\lbrace{+,-\rbrace}^n$ and $V \in \mathbb{B}_n^X$. We have $V \ketbra{\s}{\s} V^\dagger = \ketbra{\Phi(\s,V)}{\Phi(\s,V)}$ for some pure state $\ket{\Phi(\s,V)}$. We also have $\ketbra{\s}{\s}=\sum_{\alpha} c_{\alpha} X_{\alpha}$ for some family of complex coefficients $\{c_{\alpha}\}$. Here we recall our notation where $\alpha$ is a bit-string of $n$ bits, where the bit $1$ (resp $0$) at the $i$'th position indicates the $X$-Pauli (resp $I$) should be applied on the $i$'th tensor product, i.e.  $X_\alpha =\prod_{i\in\alpha} X_i$. Now, because $V X_{\alpha} V^{\dagger} \in \mathbb{U}_n^X$, we deduce that $\ketbra{\Phi(\s,V)}{\Phi(\s,V)}$ is diagonal in the local $X$-basis, i.e., the basis composed of the elements $\{\ket{\s}, \s \in \lbrace{+,-\rbrace}^n \}$. Being a pure state, it implies   $\ket{\Phi(s,V)} \propto \ket{\s'}$ for some $\s'\in \lbrace{+,-\rbrace}^n$. Because unitary channels are invertible we see that we can define a permutation $\sigma_V: \lbrace{+,-\rbrace}^n \rightarrow \lbrace{+,-\rbrace}^n$ such that for all $\s\in\lbrace{+,-\rbrace}^n$ we have $V \ketbra{\s}{\s} V^\dagger = \ketbra{\sigma_V(\s)}{\sigma_V(\s)}$. This equation implies:
\begin{equation}
V\ket{\s} = \exp(i\varphi_{\s,V}) \ket{\sigma_V(\s)}\,
\label{app:eq:Vapp}
\end{equation}
where $\varphi_{\s,V}$ is a real phase that can depend on $\s$ and $V$. 

Reciprocally, if for any $\s\in \lbrace{+,-\rbrace}^n$, there exists a real phase $\varphi_{\s,V}$ such that: $V \ket{\s}=e^{i \varphi_{\s,V}}\ket{\sigma_V(\s)}$, we have:
\begin{align}
    \sum_\s V c_\s \ketbra{\s}{\s}V^{\dagger}= \sum_\s c_\s \ketbra{\sigma_V(\s)}{\sigma_V(\s)} \in \mathbb{U}_n^X.
\end{align} 
Hence, any operator diagonal in the local $X$-basis remains diagonal in this basis through the application of the map $U \to V U V^{\dagger}$. It implies that any $V$ satisfying \eqref{app:eq:Vapp} is necessarily bias-preserving.
\end{proof}

\subsection{Coherently controlled bias-preserving gates limiting the propagation of errors}
\label{app:controlled_unitaries}
\begin{figure}[h!]
    \centering
    \includegraphics[width=\columnwidth]{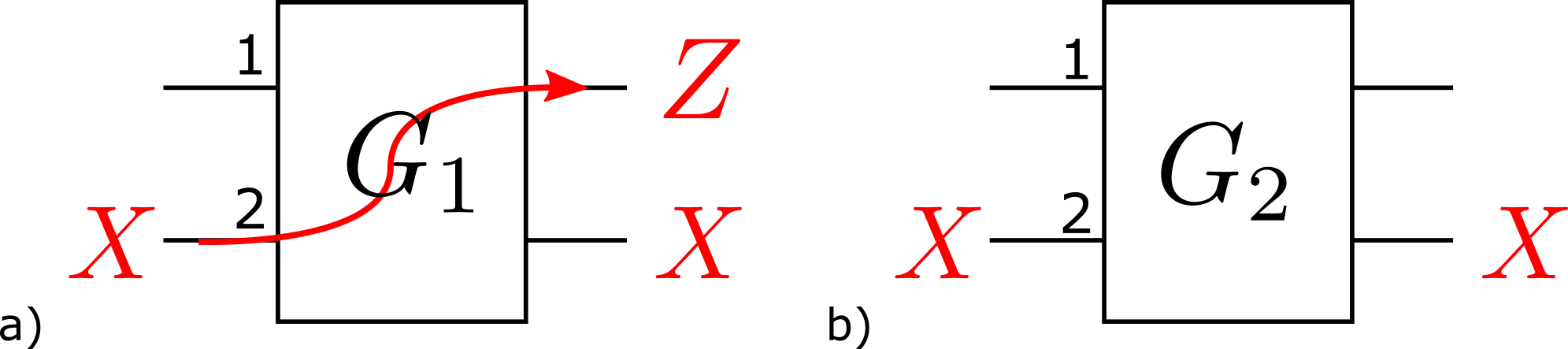}
    \caption{\textbf{a)}: A pre-existing $X$-Pauli error on qubit 2 before some two-qubit gate $G_1$ remains on qubit 2, but introduces a "new" error on qubit 1 because of the way such error propagates through $G_1$. What is shown graphically here is simply the equality: $Z_1 X_2 = G_1 X_2 G_1^{\dagger}$. \textbf{b)} In some cases, the $X$ errors do not propagate and stay on the same qubit, as shown here. Concrete examples of how errors propagate through different gates are shown in Figure \ref{fig:bias_preserving_cNOT_Toffoli}.}.
\label{fig:error_propagation_definition}
\end{figure}

While we need to preserve $X$ errors along the computation, we also need to understand how such errors propagate in order to avoid the measurement being corrupted by too many errors. What we call "error propagation" is the property on which a pre-existing Pauli error occuring on some of the qubits before a gate will introduce errors on potentially additional qubits after the gate. These "new" errors are a consequence of the gate dynamics (they would be introduced even if the gate was noiseless). An illustration of what we mean is shown in Figure \ref{fig:error_propagation_definition}.
Because for the Hadamard test, the measured qubit only interacts with other qubits under controlled-unitary operations (the measured qubit being on the control side), and because the measured qubit is the one that has to be isolated from the noise, we focus our interest on the conditions in which $X$ errors do not propagate from the target to the control side of the unitary. We call $c_{P} U$ a coherently controlled unitary $U$ applied if the control qubit is the eigenstate $-1$ of the single-qubit matrix $P \equiv \boldsymbol{n}.\boldsymbol{\sigma}$, where $\mathbf{n}$ is a unit vector, and $\boldsymbol{\sigma}=(X,Y,Z)$. The first thing we can easily notice is that $U$ must commute with \textit{any} element in $\mathbb{P}_n^X$. Otherwise, for some state $\ket{\psi}$, for some $P_X \in \mathbb{P}_n^X$, we would have:
\begin{align}
\bra{\psi} P_X^{\dagger} U P_X \ket{\psi} \neq \bra{\psi} U \ket{\psi}
\label{app:eq:app:A_commutes_X}
\end{align}
It would imply that there would exist a Hadamard test, implemented on the unitary $U$ and a state $\ket{\psi}$ that would be sensitive to $X$-errors occuring in the data register. Hence, $X$-errors would propagate from the data to the measured register. This remark can be summarized as follow:
\begin{property}[Necessary conditions on $U$ such that $c_P U$ does not propagate errors from the target to the control]
\label{app:prop:necessary_cond_A_in_cPA}

If the controlled unitary $c_P U$ acting on $1+n$ qubits, defined as:
\begin{align} c_P U \equiv \frac{1}{2}(I+P) \otimes I + \frac{1}{2}(I-P) \otimes U,
\end{align}
does not propagate $X$ errors from the target to the control, then $U$ commutes with any element in $\mathbb{P}_n^X$, which implies $U \in \mathbb{U}_n^X$.

Here, $P \equiv \mathbf{n}.\boldsymbol{\sigma}$, where $\mathbf{n}$ is a unit vector, and $\boldsymbol{\sigma}=(X,Y,Z)$. The absence of propagation of $X$-errors from the target to the control is defined as:
\begin{align}
\forall P_X \in \mathbb{P}_n^X, (c_P U) (I \otimes P_X) (c_P U)^{\dagger}=I \otimes B
\end{align}
for some unitary $B$.
\end{property}
What we need to know is whether or not there are additional constraints on $U$ than being in $\mathbb{U}_n^X$. It is the case, depending on the choice of $P$. Here, we give the definition of a controlled-unitary between the measured and data register which is allowed in the computation: it shouldn't propagate $X$-errors from the target to the control, and it should also preserve the noise bias.
\begin{definition}[Bias-preserving controlled unitaries avoiding $X$-errors to propagate toward the control]
\label{def:app:bias_preserving_controlled_unitaries}

We say that the controlled-unitary 
$$c_{P} U \equiv \frac{1}{2}(I+P) \otimes I + \frac{1}{2}(I-P) \otimes U$$ acting on $1+n$ qubits is bias preserving, and does not propagate $X$-errors to the control, if it satisfies
\begin{align}
    &\forall P_X \in \mathbb{P}_{n+1}^X, (c_{P} U) P_X (c_{P} U)^{\dagger} \in \mathbb{U}_n^X \label{C1}\\
    &\forall P_X \in \mathbb{P}_n^X, \exists U_X \in \mathbb{U}_n^X \text{ such that,} \notag \\
    &(c_{P} U) \left(I \otimes P_X\right)(c_{P} U)^{\dagger} = (I \otimes U_X) \label{C2},
\end{align}
\end{definition}
The following property characterises the controlled unitaries between the measured and data register which are allowed in the computation (hence the ones which do not propagate $X$-errors from target to control, and preserve the noise bias).
\begin{property}[Characterisation of controlled unitaries avoiding errors to propagate toward the control]
\label{app:prop:bias_preserving_controlled_unitariesZX}

The controlled unitary $c_{X} U$ is biased preserving and avoids $X$ errors to propagate toward the control if and only if $U \in \mathbb{U}_n^X$.

For $(y,z)$ being a 2D unit vector ($y,z \in \mathbb{R}$), the controlled unitary $c_{P} U$ with $P=yY+zZ$ is biased preserving and avoids $X$ errors to propagate toward the control if and only if $U \in \mathbb{U}_n^X$ and $U$ is Hermitian.

There doesn't exist any controlled unitary $c_{P} U$ with any other $P=\boldsymbol{n}.\boldsymbol{\sigma}$ than previously discussed (where $\mathbf{n}$ is a unit vector, and $\boldsymbol{\sigma}=(X,Y,Z)$) that is not trivial (i.e. $U \neq I$) and that will satisfy the constraints on error propagation.
\end{property}
\begin{proof}
Now, we show Property \ref{app:prop:bias_preserving_controlled_unitariesZX}. We need $c_{P}U$ to satisfy the conditions \eqref{C1} and \eqref{C2}. We begin with \eqref{C2}:
\begin{align}
   c_P U (I \otimes P_X) (c_P U)^{\dagger}=I\otimes U_{X}
\end{align}
for some $U_{X} \in \mathbb{U}_{n}^{X}$ and for any $P_X \in \mathbb{P}_{n}^{X}$. This equation implies:
\begin{align}
    (I+\boldsymbol{n}\cdot \boldsymbol{\sigma})\otimes P_{X}+(I-\boldsymbol{n}\cdot \boldsymbol{\sigma})\otimes UP_{X}U^{\dagger}=2I\otimes U_{X}. \label{C2-Simplified}
\end{align}
Expanding it in the Pauli basis and identifying the left and right hand sides, \eqref{C2-Simplified} implies that for any $P_X \in \mathbb{P}_n^X$, $P_X=UP_{X}U^{\dagger}$ and $P_{X}+UP_{X}U^{\dagger} = 2 U_X$. These equations are not independent and the first one implies the second one. As we already know that $U \in \mathbb{U}_n^X$ (see property \ref{app:prop:necessary_cond_A_in_cPA}), we realize that no additional constraints on $U$ (rather than $U \in \mathbb{U}_n^X$) are found here. Hence, we move on with the condition \eqref{C1}.

The condition \eqref{C1} implies that for all $P_X \in \mathbb{P}_n^X$, $(c_PU) I \otimes P_{X}(c_{P}U)^{\dagger} \in \mathbb{U}_{n+1}^X$ and $(c_PU)X\otimes P_{X}(c_{P}U)^{\dagger} \in \mathbb{U}_{n+1}^X$. The first condition is already implied by \eqref{C2} that we just analyzed, hence we focus on the second one. By considering $\boldsymbol{n}=(x,y,z)$, and using the fact $x^2+y^2+z^2=1$, we find that $(c_PU)X\otimes P_{X}(c_{P}U)^{\dagger} \in \mathbb{U}_{n+1}^X$ implies:
\begin{align}
\frac{x}{2} I\otimes\big(P_{X}-UP_{X}U^{\dagger}\big)\notag\\
  +\frac{1}{2}X\otimes \big(x^2 (P_{X}+UP_{X}U^{\dagger})+(y^2+z^2) (P_{X}U^{\dagger}+UP_{X})\big) \notag\\+\frac{1}{2}Y'\otimes\big(UP_{X}-P_{X}U^{\dagger}\big)
    \notag\\+\frac{1}{2}Z'\otimes \big(P_{X}-P_{X}U^{\dagger}-UP_{X}+UP_{X}U^{\dagger}\big) \in \mathbb{U}_{n+1}^X.
    \label{Expand-C1}
\end{align}
where $Y'\equiv iyZ-izY$ and $Z'=x(yY+zZ)$. One can easily check that the matrices in the set $\{I,X,Y',Z'\}$ form a basis for the space of $2\times 2$ complex matrices.
From this equation, knowing that $U \in \mathbb{U}_n^X$, the only non-trivial implications are: 
\begin{align}
Y'\otimes\big(UP_{X}-P_{X}U^{\dagger}\big)=0 \label{Y'-Component}\\
   Z'\otimes \big(P_{X}-P_{X}U^{\dagger}-UP_{X}+UP_{X}U^{\dagger}\big)=0 \label{Z'-Component}
\end{align}
\textbf{First case: either $y\neq 0$ or $z\neq 0$:}\\
Knowing that $U \in \mathbb{U}_n^X$, \eqref{Y'-Component} implies $U=U^{\dagger}$. Then, \eqref{Z'-Component} implies that we must either have $x(yY+zZ)=0$, or $P_{X}-P_{X}U^{\dagger}-UP_{X}+UP_{X}U^{\dagger}=0$. If $P_{X}-P_{X}U^{\dagger}-UP_{X}+UP_{X}U^{\dagger}=0$ using the fact that $U$ is Hermitian and belongs to $\mathbb{U}_{n}^X$, the only solution for $U$ is $U=I$ which is a trivial gate. Hence, in order to find non-trivial gates, we must have $x=0$. \textit{In conclusion, \eqref{Y'-Component}, \eqref{Z'-Component} imply that if $y\neq 0$ or $z \neq 0$ then necessarily $U$ is Hermitian. Additionally, we must have $x=0$ so that $U \neq I$}.\\
\textbf{Second case: $y=z=0 \Leftrightarrow x=1$:}\\
In such a case, there are no additional constraints: any $U \in \mathbb{U}_n^X$ will satisfy the constraints \eqref{Y'-Component}, \eqref{Z'-Component}.

Here, we provided necessary conditions on $U$, $(x,y,z)$ but one could check that injecting any of the solutions would satisfy the constraints. Hence our conditions are necessary and sufficient.
\end{proof}

\subsection{Example of bias-preserving gates}
\label{app:example_bias_pres}
Here, we provide concrete examples of bias-preserving gates and we show how they propagate errors in circuits. It is worth noticing that while bias-preserving gates do exist "in principle", being able to actually implement them in a bias-preserving manner is not straightforward in general. This is because, in the laboratory, these gates are implemented through continuous Hamiltonian evolution. If the Hamiltonian used to perform the gate contains Pauli $Z$-terms, it might be that a bit-flip ($X$-error) produced \textit{during} the gate evolution will be converted to a phase flip ($Z$-error). For instance, while a cNOT preserves $X$-errors "in principle", it is not straightforward to guarantee that this condition still holds if we consider a continuous-time evolution to implement the gate. In practice, this issue can be resolved with cat-qubits \cite{Puri2020Aug}, and both the $\text{Toffoli'} \equiv H_1 H_2 H_3 \times \text{Toffoli} \times (H_1 H_2 H_3)^{\dagger}$ and cNOT can be shown to preserve $X$-errors \cite{Guillaud2021Apr,Puri2020Aug}. Here, we recall that in our work we took a different convention for the dominant source of errors (in \cite{Guillaud2021Apr,Puri2020Aug} they consider that the main source of errors are phase-flips, i.e. $Z$ errors while we consider bit-flips, i.e. $X$ errors). Hence we "convert" their results to our convention simply by "swapping" the role of $X$ and $Z$ for the gate definitions. This is why we can say that Toffoli' and cNOT preserve the bit-flip bias.

\begin{figure}[h!]
    \centering
    \includegraphics[width=\columnwidth]{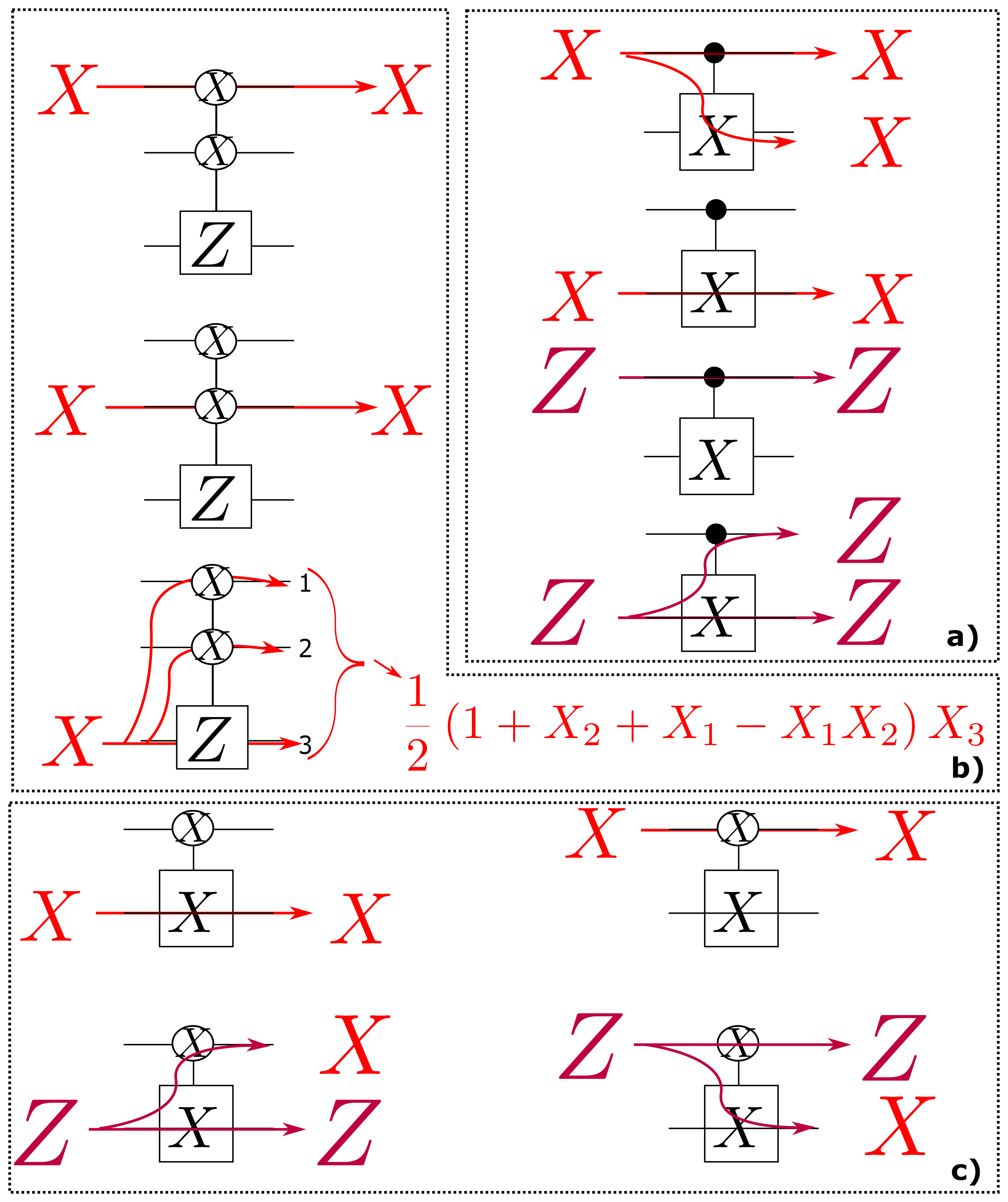}
    \caption{On this image, we represent how some initial Pauli errors ($X$ or $Z$) propagate through different bias-preserving gates. In particular, we give examples of bias-preserving gates (preserving $X$-errors) with a cNOT, Toffoli' (we call Toffoli' a Toffoli where the role of $X$ and $Z$ have been exchanged), and $c_X X$ ($c_X G$ means the application of a gate $G$ coherently controlled in the $X$ basis, i.e., applied for the control being in $\ket{-}$ and not applied if the control is in $\ket{+}$). Coherent control in the $X$ basis is represented by an $X$ inside of a black circle (visible on the top qubit on c) for instance).  On a), we see that an $X$ error in a cNOT propagates from the control to the target, but not the other way around. A $Z$ error propagates from the target to the control, but never from the control to the target. On b), we see that a Toffoli' propagates $X$-errors toward multiple qubits only in the case a pre-existing error on the target qubit occurred. In such a case, the resulting error is no longer an element of $\mathbb{P}_n^X$, but it belongs to $\mathbb{U}_n^X$. Toffoli' would not preserve $Z$ (or $Y$) errors. On c), we see that an $X$ error trivially commutes with $c_X X$. For such a gate, the $Z$ errors are not preserved: a Pauli $Z$ before $c_X X$ would be equivalent to $c_X X$ followed by an operator that is no longer a Pauli $Z$ operator as shown in the two examples at the bottom.}
\label{fig:bias_preserving_cNOT_Toffoli}
\end{figure}

The propagation of errors through cNOT and Toffoli' is shown in figure \ref{fig:bias_preserving_cNOT_Toffoli}. There, we see that depending on which qubit had a pre-existing $X$-Pauli error, this error can propagate toward multiple qubit. Sometimes, the resulting error is no longer a Pauli operator as shown in figure \ref{fig:bias_preserving_cNOT_Toffoli} b). However, because we consider bias-preserving gates, we are guaranteed that for pre-existing Pauli-X errors, the resulting error after the gate is an element of $\mathbb{U}_n^X$. In this figure, we also showed how $Z$ errors would propagate through a cNOT (the definitions provided in the main text can naturally be generalized to $Y$ and $Z$ operators). The propagation for $Y$ errors can be deduced from the propagation of $Y$ and $Z$ errors. Toffoli' does not preserve $Z$ errors (a pre-existing $Z$ error on any of the control would not be anymore a $Z$ error after the gate).

\section{Properties of the preparable states $\ket{\psi}$}
\subsection{Diagonal unitaries in $X$ basis for the preparation are unnecessary}
\begin{property}
\label{app:prop:diagonal_X_basis_useless_prep}

We consider the preparation unitary of the Hadamard test, $B$, to be a product of bias-preserving gates. We have the following results:
(i) any gate in $\mathbb{U}_n^X$, used inside the unitary $B$ cannot change the expectation value $\bra{\psi} U \ket{\psi}$ for $U \in \mathbb{U}_n^X$. (ii) Considering the noise model from definition \textcolor{Blue}{4} of the main text is satisfied, any gate in $\mathbb{U}_n^X$, used inside the unitary $B$ cannot change the measurement outcomes of the circuits we use in our work (i.e. the ones satisfying the constraints of Theorem \ref{app:thm:noise_resilience}).
\end{property}
\begin{proof}
We first show (i). In order to show this property, we will first assume that $B=C \times B'$, where $C \in \mathbb{U}_n^X$. We then have:
\begin{align}
    \ket{\psi} \equiv C \ket{\Psi} = C B' \bigotimes_{i=1}^n \ket{\phi_i},
\end{align}
where the family $\{\ket{\phi_i}\}_{i=1}^n$ represents the initial single qubit states prepared in the data register. In such a case, we have:
\begin{align}
    \bra{\psi} U \ket{\psi} = \bra{\Psi}C^{\dagger} U C \ket{\Psi}=\bra{\Psi}U\ket{\Psi},
\end{align}
where we used the fact that $[C,U]=0$ for the last equality (because $(C,U) \in \mathbb{U}_n^X$). Hence, it shows that there is no interest in applying a unitary $C \in \mathbb{U}_n^X$ right before the controlled $U$ as it won't have any influence on the expectation value. It remains to understand what happens if $C$ was applied before this point. Let's assume for instance that $B=B_1 C B_2$, where $C \in \mathbb{U}_n^X$. Using the fact that all the gates used in $B$ belong to $\mathbb{B}_n$, $B_1$ is also in $\mathbb{B}_n$. Hence, $B_1 C = C' B_1$, with $C' \in \mathbb{U}_n^X$. As shown above, $C'$ cannot change the expectation value $\bra{\psi} U \ket{\psi}$. Hence, it means that implementing $B=B_1 C B_2$, or implementing instead $B_1 B_2$ would lead to the same result. Overall, we showed that any gate $C \in \mathbb{U}_n^X$ used in the preparation unitary $B$ can be removed as it doesn't change the expectation value for the Hadamard test. Finally, we show (ii). Consider first the circuit to be noiseless. Assume we implement a unitary $G \in \mathbb{U}_n^X$ inside $B$, i.e. assume $B=B_1 G B_2$, with $(B_1,B_2) \in \mathbb{B}_n$. We can "commute" $G$ toward the end of the preparation: we have $B=G' B_1 B_2$ with $G' = B_1 G B_1^{\dagger} \in \mathbb{U}_n^X$ (because $B_1 \in \mathbb{B}_n$). Because, after the implementation of $B$, any gate applied on the data register commutes with any element in $\mathbb{U}_n^X$ locally applied on the data register (see the assumptions we use for our circuits in Theorem \ref{app:thm:noise_resilience} and Figure \ref{fig:parallelisation_register}), $G'$ cannot modify the measurement outcomes of the circuit. Now, if the circuit is noisy, with the noise model satisfying definition \textcolor{Blue}{4} of the main text, the same conclusion would hold. This is because the Kraus operators of every gates are a linear combination of operators belonging to $\mathbb{U}_n^X$: a similar reasonning (i.e. commuting the now noisy implementation of $G$ toward the end of the preparation), would also show that the noisy implementation of $G$ can be completely removed from the circuit without impacting the measurement outcomes. More precisely, in this case, one can show that the quantum channel associated to the preparation unitary $B$, $\mathcal{E}_B$, is equal to $\mathcal{E}_B=\mathcal{A} \circ \mathcal{G}_{B_1} \circ \mathcal{G}_{B_2}$ with $\mathcal{A}$ being a quantum channel having Kraus operator that are linear combination of operators in $\mathbb{U}_n^X$, and $\mathcal{G}_{B_i}(\rho)=B_i \rho B_i^{\dagger}$ ($i \in \{1,2\}$). Thus, $\mathcal{A}$ will commute with any noisy gate applied after the preparation, in the circuit of Figure \ref{fig:parallelisation_register}, as they commute with any element in $\mathbb{U}_n^X$ locally applied on the data register.
\end{proof}
Alternatively, Property \ref{app:prop:diagonal_X_basis_useless_prep} also implies that all the single qubit states $\ket{\phi_i}$ that differ by a rotation around the $X$-axis of the Bloch sphere would lead to the same expectation value (for instance initializing $\ket{\phi_i}$ in an eigenstate of $Y$ or $Z$ would provide the same outcome for the Hadamard test).
\subsection{Additional remarks about entanglement properties}
\label{app:add_prop_entanglement}
Here, we wish to comment on different properties about the kind of entanglement that can be generated with bias-preserving circuits, in general, and how it can be used in our specific examples based on the Hadamard test. In the last paragraph of the discussion part of the main text, we explained that highly entangled graph states can be generated. This is done by initializing the data register in $\ket{0}^{\otimes n}$, and applying some gates $c_X X \in \mathbb{U}_n^X$ in the preparation unitary $B$ \cite{hein2004multiparty}. It shows that, in general, bias-preserving circuit can generate quantum states having interesting computational properties, a remark worth noticing for possible extensions of our work. 

However, it should be noticed that, as a consequence of Property \ref{app:prop:diagonal_X_basis_useless_prep}, these $c_X X$ gates cannot modify the measurement outcomes of the circuits we are using (hence they can be removed without impacting experimental outcomes).
Note that this is not an issue for our goals: (i) even if $B$ is exclusively composed of gates in $\mathbb{U}_n^X$, the benchmarking would be useful (ii) there exists entangled states for which the entangling operations used in their preparation modifies the measurement outcomes of the circuits. For (i), the purpose of the benchmarking is to check whether or not the circuit works as expected, i.e. that it has a perfectly biased noise model. As long as the bias is \textit{imperfect} (i.e. whenever it does not satisfy definition \textcolor{Blue}{4} of the main text exactly), the gates inside $B$ will generate errors that \textit{will}, in general, impact the measurement outcomes (even if all gates in $B$ belong to $\mathbb{U}_n^X$). Thus, implementing the benchmarking on such circuits will in general allow to detect imperfection of the bias (because a perfect and imperfect bias would in general lead to different measurement outcomes). This is an information of experimental relevance: it means, in particular, that we can detect imperfection in the bias for a circuit preparing graph states.

For (ii), a simple example can be an $n$-qubit GHZ state. It can be created by initializing the first qubit of the data register in $\ket{+}$, the rest of the qubits in $\ket{0}^{\otimes {n-1}}$ and by applying a sequence of cNOTs controlled by the first qubit (targetted on the rest of the qubits). While belonging to $\mathbb{B}_n^X$, the cNOTs do not belong to $\mathbb{U}_n^X$. Because of that, the preparation unitary $B$ itself will, in general, have an influence on the measured outcome\footnote{For instance, a Hadamard test with $U=X_1$ will satisfy $\bra{\psi}U\ket{\psi}=1$ for $\ket{\psi}=\ket{+}\ket{0}^{\otimes {n-1}}$, but $\bra{\psi}U\ket{\psi}=0$ for $\ket{\psi}=\ket{GHZ}$, with $\ket{GHZ}$ the GHZ state mentionned: the cNOTs will thus modify the measurement outcomes of our circuits, in general.}.

\section{Proof of the results assuming Pauli bit-flip noise}
\subsection{Preliminary result}
\setcounter{property}{0}
\begin{property}[Preservation of the bias]
\label{prop:bias_preservation}
If a quantum circuit is only composed of gates in $\mathbb{B}_n$, each subject to the local biased noise model (\textcolor{Blue}{2}) of the main text (and the paragraph that follows for measurement and preparation), then any error affecting the state of the computation is an $X$-error.
\end{property}
\setcounter{property}{5}
\begin{proof}
    After initialization, in the absence of errors, the state of the computation would be $\ket{\psi_{\text{prep}}}=\bigotimes_{i=1}^n \ket{\phi_i}$. Because the initialization is noisy, following the noise model described around (\textcolor{Blue}{2})  (of the main text), the state after preparation will be in some mixture: 
    \begin{align}
        \rho_i = \sum_{\alpha} p^{\text{prep}}_\alpha X_{\alpha} \ketbra{\psi_{\text{prep}}}{\psi_{\text{prep}}} X_{\alpha}^{\dagger}
        \label{app:eq:app:post_prep}
    \end{align}
    for some probabilities $p^{\text{prep}}_{\alpha}$. In this equation (and the equations that follow), the sum over $\alpha$ is such that all $n$-qubit $X$-Pauli operator will be reached exactly once, for some $X_{\alpha}$ (more formally, the sum is such that $\alpha \subset \mathrm{supp}(I_{n})$ for $I_n$ an identity matrix applied on $n$ qubits). We now consider a gate $G \in \mathbb{B}_n$. For any state $\ket{\Psi}$, we have, for some probability $p^G_{\alpha}$: 
    \begin{align}
    \mathcal{E}(\ketbra{\Psi}{\Psi})&=\mathcal{N}_{\mathcal{G}} \circ \mathcal{G} (\ketbra{\Psi}{\Psi}) \notag \\
    &=\sum_{\alpha} p^G_{\alpha} X_{\alpha} G \ketbra{\Psi}{\Psi} G^{\dagger} X_{\alpha}^{\dagger}
    \label{app:eq:app:post_G}
    \end{align}
    Hence, we have:
    \begin{align}
        \mathcal{E}(\rho_i)=\sum_{\alpha_1,\alpha_2} p^G_{\alpha_1} p^{\text{prep}}_{\alpha_2} X_{\alpha_1} G  X_{\alpha_2} \ketbra{\psi_{\text{prep}}}{\psi_{\text{prep}}} X_{\alpha_2}^{\dagger} G^{\dagger} X_{\alpha_1}^{\dagger}
    \end{align}
    Using the fact $G \in \mathbb{B}_n$, we have: $G  X_{\alpha_2}=E_{\alpha_2} G$ for some $E_{\alpha_2} \in \mathbb{U}_n^X$. Hence:
    \begin{align}
        \mathcal{E}(\rho_i)=\sum_{\alpha_1, \alpha_2} p^G_{\alpha_1} p^{\text{prep}}_{\alpha_2} X_{\alpha_1}   E_{\alpha_2} G \ketbra{\psi_{\text{prep}}}{\psi_{\text{prep}}} G^{\dagger} E_{\alpha_2}^{\dagger}  X_{\alpha_1}^{\dagger}
    \end{align}
    Because $X_{\alpha_1} \times E_{\alpha_2} \in \mathbb{U}_n^X$, it shows that the noiseless and noisy state of the computation (after noisy preparation and noisy gate $G$) only differ through the presence of errors in $\mathbb{U}_n^X$. The same reasoning could be applied recursively for any gate applied after this point, showing that the property \ref{prop:bias_preservation} is true.
    \end{proof}
\subsection{Proof of Theorem I of the main text (with noiseless identity gates)}
\begin{figure}[h!]
    \centering
\includegraphics[width=\columnwidth]{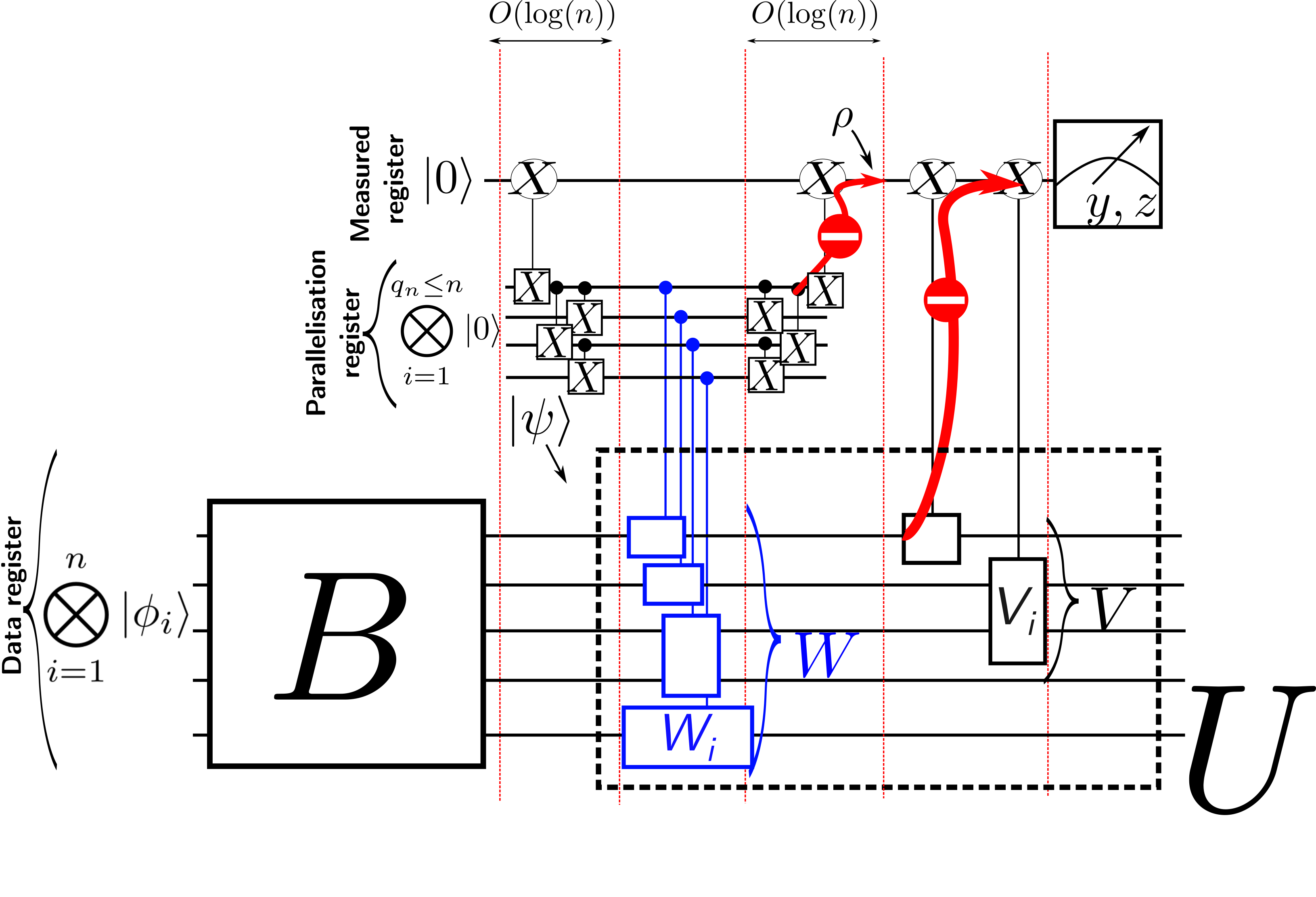}
    \caption{In this figure, we illustrate how $U=W \times V = \bigotimes_{i=1}^{N_W} W_i \times \bigotimes_{i=1}^{N_V} V_i$ mentionned in theorem \ref{app:thm:noise_resilience} can be implemented in a noise-resilient manner. The coherently controlled unitaries with a blue contour implement the controlled Hermitian unitaries $W_i$ which overall implement the controlled $W$. The coherently controlled unitaries with a black contour implement the controlled $V_i$, which overall implement the controlled $V$. The central element in this construction is the parallelisation register \cite{moore2001parallel}, which allows to implement the unitary $W$ in such a way that it acts upon all the qubits in the data register while preserving the noise-resilience. Hence, it allows to implement a Hadamard test with $U$ acting on all the qubits of the data register. Such unitaries are useful for our benchmarking protocol: see the third paragraph in the discussion section. The key point behind the parallelisation register is that, while it introduces additional components that can introduce bit-flip errors, these bit-flips cannot, by construction, propagate to the measured register (in practice, they commute with the last $c_X X$ gate drawn). It is an example where trading space (using more qubits) to gain time (guaranteeing that the measured register interacts with $O(\log(n))$ gates and not $\text{poly}(n)$ gates) is worth doing. While the parallelisation register does not propagate bit-flips toward the measured register, it will propagate phase-flip errors. We make use of this to easily detect an excessive production of phase-flip errors in the third paragraph of the discussion section of the main text.}
\label{fig:parallelisation_register}
\end{figure}

\begin{restatable}[Hadamard test resilient to biased noise]{theorem}{HadamResNoise}
\label{app:thm:noise_resilience}
Let:
\begin{align}
& \ket{\psi}=B \bigotimes_{i=1}^{N_B} \ket{\phi_i}, \ U=W \cdot V\ , \notag \\
&\ W \equiv \prod_{i=1}^{N_W} W_i, \ V \equiv \prod_{i=1}^{N_V} V_i\ ,
\label{eq:U_main_result}
\end{align}
where $B$ is a product of local bias preserving gates, gates $V_i$ and $W_i$ are local gates and belong to $\mathbb{U}_n^X$. Additionally, the gates $W_i$ are assumed to be Hermitian. We assume the circuit is implemented as indicated on Figure \ref{fig:parallelisation_register}. There, $W$ is implemented thanks to the "parallelisation register", while $V$ is implemented by making the measured and data register directly interact.

Furthermore, we assume the local bias noise model introduced in Eq. (\textcolor{Blue}{2})  of the main text, and that state preparation, measurements, and each non-trivial gate applied on the measurement register have a probability at most $p<1/2$ to introduce a bit-flip on the measured register.

Under these conditions, there exists a quantum circuit realising a Hadamard test such that, in the presence of noise, the reduced state $\rho$ satisfies Eq.~(\textcolor{Blue}{1}) with $\alpha_n \geq (1-2 p)^{O(N_V)}$.

Additionally, $\alpha_n$ is efficiently computable classically. Hence, if $N_V=O(\log(n))$, it is possible to implement the Hadamard test in such a way that running the algorithm $\text{poly}(n)$ times is sufficient to estimate the real and imaginary parts of $\bra{\psi} U \ket{\psi}$ to $\epsilon$ precision with high probability.
\end{restatable}
\begin{proof}
We focus our interest on figure \ref{fig:parallelisation_register}. In this image, the controlled unitaries with a blue contour correspond to the $W_i$, and the ones with a black contour to the $V_i$. We first explain how the $U$ defined in Theorem \ref{app:thm:noise_resilience} is implemented. Then we will show that the circuit is noise-resilient. Attributing the indices $m,p,d$ for, respectively, the measured, parallelisation and data registers, and calling $q_n \leq n$ the number of qubits in the parallelisation register, the full state of the computer, in the noiseless case, right before the application of any controlled $W_i$ or $V_i$ is the entangled state:
\begin{align}
   &\ket{\Psi}_0 \equiv \frac{\ket{+}_{m}\ket{0}^{\otimes q_n}_{p} +\ket{-}_{m} \ket{1}^{\otimes q_n}_{p} }{\sqrt{2}} \otimes \ket{\psi}_{d}.
\end{align}
This entanglement has been created thanks to a sequence of cNOT and $c_X X$ gates (see Figure \ref{fig:parallelisation_register}). Now, we apply the controlled $W_i$ and the state becomes:
\begin{align}
& \ket{\Psi}_0 \to \notag \\
&\frac{\ket{+}_{m}\ket{0}^{\otimes q_n}_{p}\ket{\psi}_{d} +\ket{-}_{m} \ket{1}^{\otimes q_n}_{p} \prod_{i=1}^{N_W} W_i \ket{\psi}_{d}}{\sqrt{2}}. 
\end{align}
The parallelisation register is then decoupled from the rest (by applying the reverse sequence of cNOT and $c_X X$). Discarding this state, and applying the last sequence of coherently controlled $V_i$, we get the required final state for the Hadamard test:
\begin{align}
& \ket{\Psi}_f = \frac{\ket{+}_{m}\ket{\psi}_{d} +\ket{-}_{m} U \ket{\psi}_{d}}{\sqrt{2}},
\end{align}
where 
\begin{align}
U=\prod_{i=1}^{N_W} W_i \times \prod_{i=1}^{N_V} V_i.
\end{align}
It proves that the appropriate operation is implemented. The parallelisation gadget we use is based on \cite{moore2001parallel}.

Now, we consider the noisy implementation of the circuit, and we show that the circuit is noise-resilient. We call $p_i$ the probability that the $i$'th gate applied on the measured register introduces a bit-flip error there. For a multi-qubit gate $\mathcal{G}_i$, considering that the measured register is the first qubit in the tensor decomposition, this probability is defined as:
\begin{align}
    p_{i} \equiv \sum_{\substack{(1,\alpha')\subset \mathrm{supp}(\mathcal{G}_i)}} p^{\mathcal{G}_i}_{1,\alpha'}
    \label{app:eq:p_G_def}
\end{align}
To clarify the notation with an example, it means that for a two-qubit gate having the noise model:
\begin{align}
    \mathcal{N}(\rho)=\sum_{\substack{0 \leq i_1 \leq 1 \\ 0 \leq i_2 \leq 1}} p_{i_1, i_2} (\sigma_{i_1} \otimes \sigma_{i_2}) \rho  (\sigma_{i_1} \otimes \sigma_{i_2}),
\end{align}
the probability to introduce a bit-flip on the measured register (first tensor product) would be $\sum_{0 \leq i_2\leq 1} p_{1, i_2}$. In what follows we will also include noisy state preparation and measurements in the derivation.

We first notice only $X$-errors can occur in the whole algorithm. This is because we use the noise model described around (\textcolor{Blue}{2})  (of the main text), and all our gates are bias-preserving (see properties \ref{prop:bias_preservation} and \ref{app:prop:bias_preserving_controlled_unitariesZX}). We also notice from this last property that the reason why the $W_i$ must be Hermitian is that they are coherently controlled on the $Z$ basis. At this point, we can additionally notice that all the gates interacting with the measured register cannot propagate $X$-error from neither the parallelisation nor the data register toward the measured register: this is another consequence of property \ref{app:prop:bias_preserving_controlled_unitariesZX}. We should emphasize on this last point: the fact we trade space for time with the parallelisation register (it introduces additional qubits in the algorithm, but it allows the measured register to interact with $O(N_V) = O(\log(n))$ and not $O(N_W) \subset \text{poly}(n)$ gates) is what will allows us to guarantee the noise-resilience of our circuit. The key point in this trade is that the errors produced on the parallelisation register cannot propagate to the measured one, because they commute with the only gate interacting with the measured register: $c_X X$. Without the use of this parallelisation register, the measured register would face errors at $O(N_W)$ additional locations, ruining the scalability for $N_W=\text{poly}(n)$. Our remarks here still hold (with one additional hypothesis), considering waiting location (identity gates) to be noisy, as discussed in the following section. Overall, here we showed that all the errors that can damage the measurements are "directly" produced on the measured register, i.e. everything behaves "as if" the only noisy qubit was the measured one.

Here comes the last part of the proof: now that we know that everything behaves "as if" the only noisy qubit was the measured one, we need to compute the effect of this noise. After the $i$'th gate interacts with the measured register, the following bit-flip channel is applied on the measured register, for some $p_i$ (and $\sigma$ is some density matrix):
\begin{align}
    \Lambda_i(\sigma)=(1-p_i) \sigma +p_i X \sigma X.
\end{align}
This noise model naturally includes the noise produced after state preparation. We can also model noisy measurements with this noise channel. This is because they are modelled as perfect ones followed by a probability to flip the measured outcome. Because we measure Pauli $Y$ and $Z$, we can equivalently model them as perfect measurements preceded with a bit-flip channel. 

Now, we use the fact the bit-flip channel commutes with every gate applied on the measured register (they are coherently controlled in the $X$-basis). Hence, calling $N_n$ the number of gates applied on the measured register\footnote{We call $N_n$ the number of gates applied on the measured register (and not $L_n$ as in the figure of the main text) because we neglect noisy identity gates here. In the main text, $L_n$ includes potential noisy waiting locations. They are taken into account in the following section.} (including state preparation and measurement), the full protocol is equivalent as performing noiseless measurements in $Y$ or $Z$ bases on the state:
\begin{align}
    &\rho = (\Lambda_{N_n} \circ ... \circ \Lambda_{1})(\rho_{\text{ideal}}),\\
    &\rho_{\text{ideal}} = \frac{1}{2}(I+y Y+z Z).
\end{align}
It is straightforward to show that:
\begin{align}
    &\rho=\frac{1}{2}(I+\alpha_n(y Y+z Z)), \notag
    \\& \alpha_n=\prod_{i=1}^{N_n}(1-2p_i).
\end{align}
In the case identity gates are noiseless, we have $N_n=N_V+2+2=O(N_V)$. The first "$+2$" corresponds to state preparation and measurement, and the last one to the implementation of the $c_X X$ gates with the parallelisation register, as shown on Figure \ref{fig:parallelisation_register}. We discuss what happens if identity gates are noisy in the following section. We notice that $\alpha_n$ is efficiently computable classically (as long as $N_n \in \text{poly}(n)$). Calling $p=\max_i p_i$, it is easy to see that $\alpha_n \geq (1-2p)^{O(N_V)}$ is true. Hence, if $N_V=O(\log(n))$, $\alpha_n$ decreases at a polynomial speed at most and is efficiently computable. Following the explanations after Eq. (\textcolor{Blue}{1}) in the main text, it implies that it is possible to implement the Hadamard test in such a way that running the algorithm $\text{poly}(n)$ times is sufficient to estimate the real and imaginary parts of $\bra{\psi} U \ket{\psi}$ to $\epsilon$ precision with a probability greater than $1-\delta$.
\end{proof}
\subsection{Extension of Theorem 1 for noisy identity gates}
\label{app:extension_thm1}
In the assumption of Theorem 1, any trivial (i.e. identity) gate applied on the measured register was assumed to be noiseless. However, the use of the parallelisation register may introduce a significant number of "waiting" locations for the measured qubit (at least for the initialization of the parallelisation register, or for the implementation of $W$). Fortunately, assuming that the depth of $W$ is in $O(\log(n))$, we could include noisy identity gates in the reasonning while not changing any of our conclusions. To keep the explanation simple, we assume that all gates in the computation (including identity) last for the same amount of time. We call $p_I$ the probability that an identity gate introduces a bit-flip, and we call $N_I$ the number of identity gates applied on the measured register. In such a case, the only thing that would change in the proof of Theorem 1 would be that:
\begin{align}
    \alpha_n=(1-2p_I)^{N_I} \prod_{i=1}^{N_n}(1-2p_i).
\end{align}
There, $\prod_{i=1}^{N_n}(1-2p_i)$ is the probability that the non-trivial gates (including preparation and measurements) introduce bit-flips. The new term $(1-2p_I)^{N_I}$ corresponds to the noise introduced by identity gates. In the case $W$ has a depth in $O(\log(n))$, we could implement the algorithm with $N_I=O(\log(n))$ (see figure \ref{fig:parallelisation_register}): the algorithm would still be noise-resilient. It would be the case as $\alpha_n$ would decrease at polynomial speed while being exactly computable. If the depth of $W$ is not in $O(\log(n))$, the identity gates might lead to an exponential decay of $\alpha_n$, ruining the scalability. It would for instance be the case if the depth of $W$ is linear in $n$ (in this case, $N_I \propto n$). One key element in this discussion is that the entanglement (and dis-entanglement) between the measured and parallelisation register can be done in $O(\log(n))$ depth \cite{moore2001parallel}, as also illustrated on the figure \ref{fig:parallelisation_register}.

In conclusion, if the depth of $W$ is in $O(\log(n))$, then the main conclusion behind theorem 1 (the fact running the algorithm $\text{poly}(n)$ times is sufficient to estimate the real and imaginary parts of $\bra{\psi} U \ket{\psi}$ to $\epsilon$-precision with high probability) would also hold with noisy identity gates.
\subsection{Efficient classical simulation}

\begin{restatable}[Efficient classical simulation of restricted Hadamard test]{theorem}{EfficientSim}
\label{thm:simulation}
Let $B\in\mathbb{B}_n, U\in\mathbb{U}^X_n$ be n qubit unitaries specified by $R_B$ and $R_U$ local qubit gates (belonging to respective classes $\mathbb{B}_n$ and $\mathbb{U}^X_n$). Let $\ket{\psi_0}=\ket{\phi_1}\ket{\phi_2}\cdots\ket{\phi_n}$ be an initial product state. Then, there exists a randomized classical algorithm $\mathcal{C}$, taking as input classical specifications of circuits defining $B$, $U$, and the initial state $\ket{\psi_0}$, that efficiently and with a high probability computes an additive approximation to $\bra{\psi_0} B^\dagger U B \ket{\psi_0}$. Specifically, we have 
\begin{equation}
    \Pr\left( |\bra{\psi_0} B^\dagger U B \ket{\psi_0} -\mathcal{C}| \leq \epsilon \right)\geq 1-\delta \ ,
\end{equation}
while the running time is $T=O\left(\frac{R_B + R_U +n}{\epsilon^2} \log(1/\delta)\right)$.
\end{restatable}
    \begin{proof}
     We first note that $B^\dagger U B \in \mathbb{U}_n^X$ is diagonal in the local $X$ basis. For this reason we can rewrite
    \begin{equation}\label{app:eq:SimplifiedHad}
        \bra{\psi_0} B^\dagger U B \ket{\psi_0} = \Tr\left( B^\dagger U B\  \rho_1 \otimes \rho_2 \otimes \ldots \otimes \rho_n \right)\ ,  
    \end{equation}
    where the $\rho_i$ are states acting on the i'th qubit and defined as the : $\rho_i=\frac{1}{2}\ketbra{\phi_i}{\phi_i}+\frac{1}{2} X\ketbra{\phi_i}{\phi_i} X$ (dephased version of $\ket{\psi_i}$ in the $X$ basis). We can rewrite each $\rho_i$ as: $\rho_i=p^+_i \ketbra{+}{+}+p^-_i \ketbra{-}{-}$, where the computation of the all the probabilities $p_i^\pm$ can be done in $O(n)$ time. Equation \eqref{app:eq:SimplifiedHad} and the structure of unitaries $B,U$ can be used to derive an efficient sampling procedure for estimating $\bra{\psi_0} B^\dagger U B \ket{\psi_0}$. To this end, we decompose
    \begin{equation}\label{app:eq:DephasedInput}
         \rho_1 \otimes \rho_2 \otimes \ldots \otimes \rho_n = \sum_{\s\in\{+,-\}^n} p_\s \ketbra{\s}{\s}\, 
    \end{equation}
    where $p_\s=\prod_{i=1}^n p_i^{s_i}$ is a  product distribution. Inserting \eqref{app:eq:DephasedInput} into right-hand side of \eqref{app:eq:SimplifiedHad} we obtain
    \begin{equation}
        \bra{\psi_0} B^\dagger U B \ket{\psi_0}= \sum_{\s\in\{+,-\}^n} p_\s \Tr\left(U B \ketbra{\s}{\s} B^\dagger\right)\ .
    \end{equation}
    The above formula can be simplified further by realising that  $B\ketbra{\s}{\s} B^\dagger =\ketbra{\sigma_B(\s)}{\sigma_B(\s)}$ (cf. Property \ref{lem:BiasPreserv}) and by utilising $U\in\mathbb{U}^X_n$ (which implies $U\ket{\s}=\lambda_U(\s)\ket{\s}$, for all $\ket{\s}$). Putting this all together we obtain
    \begin{equation}\label{app:eq:unbiased estimator}
        \bra{\psi_0} B^\dagger U B \ket{\psi_0}= \sum_{\s\in\{+,-\}^n} p_\s \lambda_U(\sigma_B(\s))\ .
    \end{equation}
    
    The above equation shows that the random variables $x_\s= \mathrm{Re}(\lambda_U(\sigma_B(\s)))  , y_\s =  \mathrm{Im}(\lambda_U(\sigma_B(\s)))$ are unbiased estimators of the real and imaginary parts of $ \bra{\psi_0} B^\dagger U B \ket{\psi_0}$ respectively. This suggests a straightforward three steps algorithm for constructing estimators of the quantities of interests: (i) generate $\s\sim \{p_\s\}$, (ii) compute $\s'=\sigma_B (\s)$, (iii) evaluate  $\lambda_U(\s')$ in order to compute $x_\s,y_\s$. From Hoeffding's inequality (both $x_\s$ and $y_\s$ take values in the interval $[-1,1]$) it follows that repeating the above procedure $O((1/\epsilon^2) \log(1/\delta))$ times and taking sample means gives $\epsilon$ -accurate estimation of $ \bra{\psi_0} B^\dagger U B \ket{\psi_0}$ with probability at least $1-\delta$. 
    
    What remains to be shown is the classical computation cost of steps (i)-(iii). Since $\s$ is distributed according to the simple product measure on $n$ bits, generation of a single sample takes $O(n)$ time. With regards to (ii) we use Property \ref{lem:BiasPreserv} and the decomposition of $B$ into a sequence of $R_B$ local gates from $\mathbb{B}_n$ - since each gate acts locally, it will induce an easily tracktable transformation of $\ket{\s}$ in only a subset $\beta$ of constant size. Consequently, the overall cost of implementing (ii) is thus $O(R_B)$. Similar argument holds for the final step (iii) - now the unitary $U$ is decomposed into a sequence of local unitaries $g_j$ from $\mathbb{U}^X_n$. Naturally, we have $U\ket{\s'} = \prod_{j=1}^{R_U} \lambda_j(\s') \ket{\s'}$, where $\lambda_j(\s')$ are defined by $g_j\ket{\s'}=\lambda_j(\s')\ket{\s'}$. Since gates $g_j\in\mathbb{U}^x_n$ are local, individual eigenvalues $\lambda_j(\s')$ can be computed in constant time and depend only on the few bit in $\s'$. Overall, the cost of this step is therefore $O(R_U)$. Combining the above estimates for runtimes of each of the steps of the protocol we get the desired result. 
    \end{proof}

\section{Additional noise resilience properties}
\subsection{The circuit is noise-resilient against various scale-dependent noise}
\label{app:noise_resilience_scale_dep_noise}
In this paper, we showed that if the noise model is Pauli bit-flip for each gate, and if the probability to have an error on the measured register it is upper bounded by some \textit{constant} probability $p<1/2$, then the hadamard test we implement is noise-resilient (see Theorem \ref{app:thm:noise_resilience}). However, in current experiments, it frequently happens that the probability of error for quantum gates grows with the number of qubits used in the algorithm, hence with the problem size, $n$ (this is sometimes called a scale-dependent noise) \cite{Fellous-Asiani2021Nov,G2023Jan}. It can for instance happen because of crosstalk \cite{Wilen2021Jun,Sarovar2019Aug}. 

An interesting feature occurring in our circuits is that they can, to some extent, be resilient against various classes of scale-dependent noise: the polynomial overhead in the algorithm repetitions can be preserved in that case. Assume that the probability to introduce a bit-flip on the measured register is upper bounded by $p_n<1/2$, for any gate applied there. We assume that the probability of errors for each gate is known. The result of Theorem \ref{app:thm:noise_resilience} guarantees that $\alpha_n \geq (1-2p_n)^{O(N_V)}$, with $\alpha_n$ efficiently computable classically. Calling $C_n$ a sufficient number of algorithms calls allowing to estimate $\bra{\psi} U \ket{\psi}$ up to $\epsilon$-precision with a probability $1-\delta$, we know from the main text that we can take: $C_n=2 \log(2/\delta)/(\alpha_n \epsilon^2)$. We would then have $C_n \leq 2 \log(2/\delta)/(\epsilon (1-2p_n)^{O(N_V)})^2$. For the same reasons as explained in the main text, the scaling in $1/(\epsilon (1-2p_n)^{O(N_V)})^2$ is also optimal.

Hence, as long as $1/((1-2p_n)^{O(N_V)})$ grows polynomially with $n$, the scalability would be preserved. This is an interesting property given the current limitations of the hardware in quantum computing, that can often lead to scale-dependent noise \cite{Fellous-Asiani2021Nov}.
\subsection{The Hadamard test is resilient against noisy-measurements}
\label{app:noisy_measurements}
As shown in property \ref{app:prop:bias_preserving_controlled_unitariesZX}, the unitary $U$ used in the Hadamard test must belong to $\mathbb{U}_n^X$. Because of that, in principle, we could simply measure the data register in the $X$ basis (and remove the measured register from the algorithm) in order to deduce $\bra{\psi} U \ket{\psi}$. In this part, we show that measuring the qubits of the data register will not be scalable, as soon as the measurements are noisy and one needs to measure $\text{poly}(n)$ measurements: the measured register is in general necessary. 

To show it, we only need one example of a loss of scalability for a polynomial number of measured qubits. Here, we assume that the state to be measured is guaranteed to be in some product state $\bigotimes_{i=1}^n \ket{\phi_i}$ ($\ket{\phi_i} \in \{\ket{+},\ket{-}\}$), unknown to the experimentalist. Its goal is to find if the state is an eigenstate $+1$ or $-1$ of the observable $X^{\otimes n}$. In order to find this out, it measures all the qubits in the $X$-basis. Unfortunately, these measurements have a probability $p_{\text{meas}}$ to be wrong (i.e. to return $-1$ while the measurement should produce $+1$, and vice-versa). 

We can easily compute the probability to obtain the correct measurement outcome after a single trial: it corresponds to the probability that an even number of qubits were wrongly measured. We have:
\begin{align}
    p_{\text{correct}}&=\sum_{i=0, i \ \text{even}}^n \binom{n}{i} p_{\text{meas}}^{i}(1-p_{\text{meas}})^{n-i} \notag \\
    &=\frac{1}{2}(1+(1-2p_{\text{meas}})^n)>1/2
\end{align}
In practice, the experimentalist will have access to estimators of $p(+1)$ and $p(-1)$: the probabilities that $X^{\otimes n}$ yields $+1$ or $-1$ as measurement outcome. Then, it needs to distinguish these two probability distributions. Calling $p(+1)$ the probability to have the outcome $+1$ in measuring $X^{\otimes n}$, we would have $p(+1)=p_{\text{correct}}$ in the case $+1$ was the correct outcome, and $p(-1)=1-p_{\text{correct}}$ otherwise. The goal is then to distinguish these two cases, which means to be able to distinguish two Bernoulli distribution of mean $1/2+(1-2p_{\text{meas}})^n/2$ and $1/2-(1-2p_{\text{meas}})^n/2$. Because these means are exponentially close to each other, one would necessarily need an exponential number of sample to distinguish them to a fixed error $\epsilon$ with probability larger than $1-\delta$ \cite{lectureNotesDistinguishingDistributions}. It means that measuring the qubits in the data register would, in general, not be scalable in the presence of noisy measurements. Here we took an example where exactly $n$ qubits have to be measured but it is easy to see that what matters is the polynomial scaling in the number of measured qubits.

\section{Proof of the results leading to the benchmarking}
All the results in this section assume noiseless measurements and state preparation. This is without loss of generality, up to a redefinition of the noise maps of the quantum gates: see the discussion before Theorem \ref{thm:simulation_extended} in the main text.
\label{supp:beyond_Pauli}
\subsection{Classical simulation of the noisy implementation of the circuit for a perfect bias, beyond the Pauli noise.}
We now state our efficient simulation theorem, which can simulate the outcome of the noisy hadamard test, assuming the noise is perfectly biased, according to the definition \textcolor{Blue}{4} of the main text. This efficient simulation is at the root of our general benchmarking protocol, able to check if the hardware experimentally behave as predicted theoretically.
\begin{theorem}{Efficient classical simulation of a noisy Hadamard test under perfect bias}
\label{thm:simulation_extended}

Let $B \in \mathbb{B}_n$, $U=W.V \in \mathbb{U}_n^X$, $N_W, N_V$ satisfying the constraints of Theorem \ref{app:thm:noise_resilience}. Let $\ket{\psi_0}=\ket{\phi_1}\ket{\phi_2}\cdots\ket{\phi_n}$ be the initial product state for the data register. If $N_V=O(\log(n))$, if the noise model of each gate is perfectly biased according to definition \textcolor{Blue}{4} of the main text, if the total number of gates in the algorithm is in $\text{Poly}(n)$, and if state preparation and measurements are noiseless (this is without loss of generality, see the comments before the Theorem \ref{thm:simulation_extended} in the main text), there exists a randomized classical algorithm $\mathcal{C}$ taking as input (I) classical specifications of the circuit implementing the Hadamard test for the chosen $(B,U,\ket{\psi_0})$, (II) the quantum channel describing the noise model of each gate used in the computation, (III) the initial state $\ket{\psi_0}$ such that $\mathcal{C}$ efficiently and with a high probability computes an additive approximation to $Tr(P_1 \rho_X)$, where $\rho_X$ is the reduced density matrix of the measured register at the end of the noisy implementation of the algorithm, and $P_1$ is a single-qubit Pauli matrix. 

Specifically, we have:
\begin{align}
    \Pr\left( |Tr(P_1 \rho_X) -\mathcal{C}| \leq \epsilon \right)\geq 1-\delta \ ,
\end{align}
while the running time is $T=O((1/\epsilon^2) \log(1/\delta) \times \text{poly}(n))$.
\end{theorem}
What this theorem shows is that the outcomes of the noisy implementation of the Hadamard test can be efficiently classically simulated.
\begin{proof}
  Our derivation uses the same overall idea as the one of Theorem \ref{thm:simulation}, excepted that it is now applied to the noisy implementation of the algorithm. We will show that for any single-qubit Pauli $P_1$, applied on the measured register, our classical algorithm can provide at a polynomial cost measurement outcome samples following the same probability distribution as the one they would have on the quantum computer. The theorem assumes the implementation of the circuit mentionned in theorem \ref{app:thm:noise_resilience}, hence the circuit used to implement the Hadamard test is the one of figure \ref{fig:parallelisation_register} (which allows to implement the most general Hadamard test we can). 

We begin this proof by showing that the measured register would also have the final density matrix $\rho_X$ if (H.1) only the gates applied on the measured register happened to be noisy, and (H.2), the single-qubit "$X$-dephasing" map $\Delta_X(\rho)=\frac{1}{2}(\rho+X\rho X)$ was applied after the initialization of every qubit in the parallelisation and data register. (H.1) and (H.2) will allow us to simplify our derivations.

The reason why we can assume (H.1) is because (i) we assume a perfect bias according to definition \textcolor{Blue}{4} of the main text, (ii) all the gates we use in the algorithm belong to $\mathbb{B}_n$\footnote{While the theorem \ref{app:thm:noise_resilience} makes explicit that the unitary $B$ is implemented with bias-preserving gates, note that every other gate also belong to $\mathbb{B}_n$ as a consequence of property \ref{app:prop:bias_preserving_controlled_unitariesZX}.}, (iii) the gates interacting between the measured and data or parallelisation registers commute with any $X$-Pauli operator (see Figure \ref{fig:parallelisation_register}). More precisely, any noisy gate locally applied on the parallelization or data register will damage the state of the qubits with Kraus operators being a linear combination of $X$-Pauli operators as a consequence of (i). Because we consider a gate locally acting on the parallelization and data register, and because we consider a local noise model (see definition \textcolor{Blue}{4} in the main text), these Kraus operators will act locally on these registers. These Kraus operators, when commuted with the following gates in the computation will remain a linear combination of $X$-Pauli operators as a consequence of (ii). Furthermore, as long as no gates between the measured and parallelisation or data register have been applied, these Kraus operators will act trivially on the measured register. Now, we assume that these Kraus operators has been commuted in the circuit up to the point a gate is applied between the measured register and the parallelisation or data register. Because these Kraus operators are a linear combination of $X$-Pauli operators, from (iii), they will commute with this gate. Hence, the Kraus operators will not damage the measured register density matrix (in different terms, they will act trivially on the measured register). Hence, any noisy gate applied locally on the parallelization or data register cannot modify the measured register density matrix. Hence, we can assume (H.1). We can assume (H.2) for a similar reason. Namely, applying the $X$-dephasing maps on every qubit in the parallelisation or data register after their initialisation is "as if" we had noisy identity gates, having $X$-dephasing maps as noise maps, at this place in the circuit. We just showed that any noisy gate on the parallelisation or data register cannot damage the measured register density matrix, as long as the noise is perfectly biased. Thus, we can assume (H.2). 

We call $\rho'_f$ the final density matrix the algorithm would have, under the assumptions (H.1) and (H.2). We just showed that we have $\rho_X=Tr_{\neq \text{Measured}} (\rho'_f)$, where $Tr_{\neq \text{Measured}}$ means that all degrees of freedom, apart the measured register, have been traced out.

Formally, $\rho'_f$ is defined through \eqref{eq:rhoPrimef}, but before proceeding, we recall our notations (see also section \textcolor{Blue}{II A} of the main text while reading what follows). $c_X G$ represents a unitary matrix $G$ coherently controlled in the $X$-basis. The quantum map associated to this matrix is labelled $c_X \mathcal{G}$. The noise map associated to $\mathcal{G}$ is written $\mathcal{N}_{\mathcal{G}}$. It is described with a family of Kraus operators, labelled by the letter $K$, following the notations of definition \textcolor{Blue}{4} of the main text. Now, we call
\begin{align*}
  \{A_{\substack{i_1, i_2 \\
       j_1 \cdots j_{N_V} \\
       }}\}_{\substack{i_1, i_2 \\
       j_1 \cdots j_{N_V} \\
       }}     
   \end{align*}
   the set of Kraus operators describing the sequence of noisy operations applied in the algorithm, according to figure \ref{fig:parallelisation_register} and under the assumptions (H.1) and (H.2). With these notations, we have:
\begin{widetext}
   \begin{align}
       \rho_X&=Tr_{\neq \text{Measured}} (\rho'_f)\\
       \rho'_{f}& \equiv \sum_{\substack{i_1, i_2 \\
       j_1 \cdots j_{N_V}}}
        \left(A_{\substack{i_1, i_2 \\
       j_1 \cdots j_{N_V}}}\right) \rho'_{\text{ini}} \left(A_{\substack{i_1, i_2 \\
       j_1 \cdots j_{N_V}}} \right)^{\dagger}  \label{eq:rhoPrimef} \\
       \rho'_{\text{ini}}& \equiv \ketbra{0}{0} \bigotimes_{i=1}^{q_n} \rho_i^{\text{par}} \bigotimes_{i=1}^{n} \rho_i^{\text{data}} \label{eq:rhoPrimeini}  \\
       &A_{\substack{i_1, i_2 \\
       j_1 \cdots j_{N_V} \\
       }} \equiv  K^{c_X \mathcal{G}_{V_{N_V}}}_{j_{N_V}} c_X G_{V_{N_V}} \cdots K^{c_X \mathcal{G}_{V_1}}_{j_{1}} c_X G_{V_1} K^{c_X X}_{i_2} c_X X (\mathbb{I}_{\text{meas}}\otimes \widetilde{B}) K^{c_X X}_{i_1} c_X X  (\mathbb{I}_{\text{meas}} \otimes \mathbb{I}_{\text{par}} \otimes B),
       \label{eq:Aij}\\
       \rho_i^{\text{par}}&\equiv \Delta_X(\ketbra{0}{0}) \label{eq:rhoipar}\\
       \rho_i^{\text{data}}&\equiv \Delta_X(\ketbra{\phi_i}{\phi_i}) \label{eq:rhoidata}
   \end{align}    
   \end{widetext}
   The state $\ketbra{0}{0}$ in \eqref{eq:rhoPrimeini} represents the initial state of the measured register, while $\rho_i^{\text{par}}$ and $\rho_i^{\text{data}}$ represent the $X$-dephased version of the initial states of the parallelized and data registers (see \eqref{eq:rhoipar}, \eqref{eq:rhoidata}). The term $\mathbb{I}_{\text{meas}} \otimes \mathbb{I}_{\text{par}} \otimes B$ in \eqref{eq:Aij} corresponds to the noiseless implementation of the preparation unitary on the data register ($\mathbb{I}_{\text{meas}}$ and $\mathbb{I}_{\text{par}}$ represent the identity operations applied on measurement and parallelisation registers). The term $K^{c_X X}_{i_1} c_X X$ corresponds to the first $c_X X$ gate applied between measured and parallelisation register, followed by the Kraus operator of its associated noise map. The term $\mathbb{I}_{\text{meas}}\otimes \widetilde{B}$ corresponds to the noiseless implementation of both the cNOTs applied locally on the parallelisation register, and the controlled operation $W$. The term $K^{c_X X}_{i_2} c_X X$ corresponds to the second noisy $c_X X$ gate applied between the measured and parallelisation register. $K^{c_X \mathcal{G}_{V_{N_V}}}_{j_{N_V}} c_X G_{V_{N_V}} \cdots K^{c_X \mathcal{G}_{V_1}}_{j_{1}} c_X G_{V_1}$ corresponds to the noisy application of the controlled operation $V$ as a sequence of $N_V$ gates.
  
Now, we will show that a classical computer can efficiently provide measurement outcomes samples (of the observable $P_1$ measured on $\rho_X$) that follow the same probability distribution as the one of the quantum computer.

 First, we can rewrite each $\rho^{\text{data}}_i$ as $\rho^{\text{data}}_i=p^{+ | \text{data}}_i \ketbra{+}{+}+p^{- | \text{data}}_i \ketbra{-}{-}$, and $\rho^{\text{par}}_i$ as $\rho^{\text{par}}_i=p^{+ | \text{par}}_i \ketbra{+}{+}+p^{- | \text{par}}_i \ketbra{-}{-}$ where the computation of the all the probabilities $p_i^{\pm | \text{data}},p_i^{\pm | \text{par}}$ can be done in $O(n)$ time. Then, we simply expand the initial state of the parallelisation and data registers in the eigenbasis of $X$-Pauli operators. We thus have:
    \begin{align}
         &\bigotimes_{i=1}^{q_n} \rho^{\text{par}}_i \bigotimes_{i=1}^{n} \rho^{\text{data}}_i = \sum_{\substack{\s_{\text{par}}\in\{+,-\}^{q_n}\\  \s_{\text{data}}\in\{+,-\}^n}} p_{\s_{\text{par}},\s_{\text{data}}} \ketbra{\s_{\text{par}}}{\s_{\text{par}}} \ketbra{\s_{\text{data}}}{\s_{\text{data}}}\, \label{app:eq:DephasedInputCoherentErrors1} \\
         &p_{\s_{\text{par}},\s_{\text{data}}}= p^{\text{par}}_{\s_{\text{par}}} \times p^{\text{data}}_{\s_{\text{data}}}.
         \label{app:eq:DephasedInputCoherentErrors2}
    \end{align}
    Here, $p^{\text{data}}_{\s_{\text{data}}}=\prod_{i=1}^n p_i^{s_{\text{data},i} | \text{data}}$, and $p^{\text{par}}_{\s_{\text{par}}}=\prod_{i=1}^{q_n} p_i^{s_{\text{par},i} | \text{par}}$ are product distributions. More precisely, $\s_{\text{data}}$ is a vector of length $n$ describing in which state ($+$ or $-$) each data qubit is in the mixture \eqref{app:eq:DephasedInputCoherentErrors1}, and $s_{\text{data},i} \in \{+,-\}$ represents the $i$'th component of this vector (the notation is similar for $\s_{\text{par}}$). Because the parallelisation register is composed of $q_n$ qubits, and because they all start in $\ket{0}$, we have $p^{\text{par}}_{\s_{\text{par}}}=1/2^{q_n}$. From this, we obtain:
    \begin{widetext}
    \begin{align}
        &Tr(P_1 \rho_X)=Tr(P_1 \rho'_f)=\sum_{\substack{\s_{\text{par}}\in\{+,-\}^{q_n}\\  \s_{\text{data}}\in\{+,-\}^n}} p_{\s_{\text{par}},\s_{\text{data}}} \times \lambda_{\substack{\s_{\text{par}},\s_{\text{data}}}} \label{eq:Trace_sampled}\\
        &\lambda_{\substack{\s_{\text{par}},\s_{\text{data}}}}=\sum_{\substack{i_1, i_2 \\
       j_1 \cdots j_{N_V}}}
       Tr\left( P_1 A_{\substack{i_1, i_2 \\
       j_1 \cdots j_{N_V}}} \ketbra{0}{0}  \ketbra{\s_{\text{par}}}{\s_{\text{par}}} \ketbra{\s_{\text{data}}}{\s_{\text{data}}}. (A_{\substack{i_1, i_2 \\
       j_1 \cdots j_{N_V}}})^{\dagger} \right)
        \label{eq:lambda_s}
    \end{align}
    \end{widetext}
    Now, the classical algorithm consists in (A) generate samples $\s \equiv \{\s_{\text{par}},\s_{\text{data}}\} \sim \{p_{\s_{\text{par}},\s_{\text{data}}}\}$ (i.e. generate samples $\{\s_{\text{par}},\s_{\text{data}}\}$ following the probability distribution given by $\{p_{\s_{\text{par}},\s_{\text{data}}}\}$), (B) compute $\lambda_{\substack{\s_{\text{par}},\s_{\text{data}}}}$. (A) + (B) will allow us to access measurement outcome samples (for the observable $P_1$) following the same probability distribution as the one of the quantum computer. From Hoeffding's inequality (the expectation values of $P_1$ takes values in $[-1,1]$), it follows that repeating the above procedure $O((1/\epsilon^2) \log(1/\delta))$ times and taking sample means gives $\epsilon$-accurate estimation of $Tr(P_1 \rho_X)$ with probability at least $1-\delta$. Our goal is now to estimate the complexity of performing the tasks (A) and (B), in order to estimate the complexity of the whole computation.
    
    Step (A) can be performed in $O(n)$ time classically because $\s$ is the product measure on $n+q_n=O(n)$ bits. We show that (B) is also efficient. For (B) we will first estimate the complexity of estimating one element in the summation (i.e. $(i_1,i_2,j_1 \cdots j_{N_V})$ are fixed), before multiplying it by the number of elements in the sum. We will perform the computation by evaluating $A_{\substack{i_1, i_2 \\
       j_1 \cdots j_{N_V}}} \ket{0,\s_{\text{par}},\s_{\text{data}}}$. This is done by first expanding $\ket{0,\s_{\text{par}},\s_{\text{data}}}$ in $(\ket{+,\s_{\text{par}},\s_{\text{data}}}+\ket{-,\s_{\text{par}},\s_{\text{data}}})/\sqrt{2}$ and evaluating independently the action of $A_{\substack{i_1, i_2 \\
       j_1 \cdots j_{N_V}}}$ on each of these two elements. Because all the elements in $A_{\substack{i_1, i_2 \\
       j_1 \cdots j_{N_V}}}$ are diagonal in the eigenbasis of $X$-Pauli operators, or belong to $\mathbb{B}_n$, and act on a finite number of qubits, this computation simply consists in applying some local permutations in the eigenbasis of $X$-Pauli operators, and multiplying the state by an overall complex number. We explain in more details how to compute $A_{\substack{i_1, i_2 \\
       j_1 \cdots j_{N_V}}} \ket{+,\s_{\text{par}},\s_{\text{data}}}$ (computing $A_{\substack{i_1, i_2 \\
       j_1 \cdots j_{N_V}}} \ket{-,\s_{\text{par}},\s_{\text{data}}}$ follows the exact same logic). It requires the following number of operations, starting from the state $\ket{+,\s_{\text{par}},\s_{\text{data}}}$. 
       \begin{enumerate}
           \item $O(R_B)$ to apply $(\mathbb{I}_{\text{meas}} \otimes \mathbb{I}_{\text{par}} \otimes B)$. This is because $B$ is a product of $R_B$ local permutations (permutations performed in the eigenvector basis of $X$-Pauli operators (see Property \ref{lem:BiasPreserv})), and these permutations are applied on a vector belonging to such basis.
           \item $O(1)$ for $K_{i_1}^{c_X X} c_X X$. This is because $K_{i_1}^{c_X X} c_X X$ acts (a) on a finite number of qubits, (b) is diagonal in the $X$-Pauli eigenvector basis, (c) acts on a state belonging to such basis.
           \item $O(R_{\widetilde{B}})$ for $(\mathbb{I}_{\text{meas}}\otimes \widetilde{B})$, $R_{\widetilde{B}}$ being the number of gates in $\widetilde{B}$. Similar reason as item 1. 
           \item $O(N_V)$ for \\ $K^{c_X \mathcal{G}_{V_{N_V}}}_{j_{N_V}} c_X G_{V_{N_V}} \cdots K^{c_X \mathcal{G}_{V_1}}_{j_{1}} c_X G_{V_1} K^{c_X X}_{i_2} c_X X$. This is because there are $O(N_V)$ operations in the product. Each of them takes $O(1)$ time to be performed (for similar reason as item 2). 
       \end{enumerate}
       Hence, the state after Step $4$ has the form:
       \begin{align}
           &A_{\substack{i_1, i_2 \\
       j_1 \cdots j_{N_V}}} \ket{+,\s_{\text{par}},\s_{\text{data}}}\notag \\
       &=e^{j \phi_+} \ket{+,\sigma_{\widetilde{B}}(\s_{\text{par}},\sigma_{B}(\s_{\text{data}}))},
              \label{eq:AappliedPlus}
       \end{align}
       for some phase $\phi_+$ introduced by each of the steps. The elements $\sigma_{\widetilde{B}}$ and $\sigma_B$ represent the permutations introduced by the Steps 1 and 3 on the parallelisation and data register qubits. Similar calculations are performed to compute $A_{\substack{i_1, i_2 \\
       j_1 \cdots j_{N_V}}} \ket{-,\s_{\text{par}},\s_{\text{data}}}$ which has a similar expression (for some other phase $\phi_-$):
       \begin{align}
           &A_{\substack{i_1, i_2 \\
       j_1 \cdots j_{N_V}}} \ket{-,\s_{\text{par}},\s_{\text{data}}}\notag \\
       &=e^{j \phi_-} \ket{-,\sigma_{\widetilde{B}}(\s_{\text{par}},\sigma_{B}(\s_{\text{data}}))}.
              \label{eq:AappliedMinus}
       \end{align}
       
       Because $P_1$ only acts non-trivially on the first qubit, and because the final state (which is the sum of \eqref{eq:AappliedMinus} and \eqref{eq:AappliedPlus}) is a product state, the evaluation of 
       \begin{align}
           Tr(P_1 A_{\substack{i_1, i_2 \\
       j_1 \cdots j_{N_V}}} \ketbra{0,\s_{\text{par}},\s_{\text{data}}}{0,\s_{\text{par}},\s_{\text{data}}}(A_{\substack{i_1, i_2 \\
       j_1 \cdots j_{N_V}}})^{\dagger})
       \end{align}
       takes $O(1)$ step (it consists in computing a single-qubit Pauli operator on a product state). Overall, the computation of one element inside of the sum \eqref{eq:lambda_s} takes $O(N_V+R_B+R_{\widetilde{B}})$ steps. Now, we need to count the total number of elements in the sum of \eqref{eq:lambda_s}. 
       
       Each index of the sum \eqref{eq:lambda_s} is used to refer to a Kraus operator of a given noise map. Because every gate acts on a finite number of qubits, the maximum number of values one index of the sum \eqref{eq:lambda_s} can take is finite. We call $j_{\max}$ this number. Thus, the total number of elements in the sum \eqref{eq:lambda_s} contains at most $O(j_{\max}^{N_V})$ elements. Hence, the total number of operations required to compute \eqref{eq:lambda_s} is at most $O(j_{\max}^{N_V}(N_V+R_B+R_{\widetilde{B}}))$. Hence, steps (A)+(B) can be done at $O(n+j_{\max}^{N_V}(N_V+R_B+R_{\widetilde{B}}))$ cost. 

       Finally, our goal is to access the expectation value $Tr(P_1 \rho_X)$ from the samples provided by the classical algorithm. From Hoeffding's inequality (the expectation values of $P_1$ takes values in $[-1,1]$), it follows that repeating the above procedure $O((1/\epsilon^2) \log(1/\delta))$ times and taking sample means gives $\epsilon$-accurate estimation of $Tr(P_1 \rho_X)$ with probability at least $1-\delta$. In conclusion, the overall complexity of the algorithm is no worse than $O((1/\epsilon^2) \log(1/\delta) \times (n+ j_{\max}^{N_V}(N_V+R_B+R_{\widetilde{B}})))$. Given the fact $N_V=O(\log(n))$, $R_B, R_{\widetilde{B}} \in \text{poly}(n)$, the overall task has complexity $O((1/\epsilon^2) \log(1/\delta) \times \text{Poly}(n))$ and is then computationally efficient.
\end{proof}
\subsection{Benchmarking protocol}
The assumption of a perfect bias, according to the definition \textcolor{Blue}{4} of the main text is an idealisation: in practice, the Kraus operators describing the noise maps of each component (gate, state preparation and measurement) will also involve Pauli operators not belonging to $\mathbb{P}_n^X$ in their linear decomposition on the Pauli basis. In the following section, we provide a criteria (property \ref{prop:diamond_perfect_bias}) allowing to estimate how good the perfect bias assumption is, and from that, estimate the maximum circuit size one could implement for which this assumption is reliable. We will make use of this criteria in our benchmarking protocol.
\subsubsection{Validity of the perfect bias approximation}
\label{supp:validity_perfect_bias}
We recall that we assume noiseless measurements and state preparation here, which is without loss of generality as indicated in the paragraph preceeding Theorem \ref{thm:simulation_extended} of the main text.
Let $\mathcal{G}$ be the quantum map describing a (noiseless) unitary gate, and $\mathcal{E}_\mathcal{G}$ be the quantum map describing its noisy implementation. We can define the noise map $\mathcal{N}_{\mathcal{G}}$ as:
\begin{align}
\mathcal{E}_{\mathcal{G}}=\mathcal{N}_{\mathcal{G}} \circ \mathcal{G}.
\end{align}
In this work, we assume that the noise map of each gate is perfectly biased (i.e. that it satisfies definition \textcolor{Blue}{4}). However, this is an approximation as the bias of the gate will not be perfect. It means that $\mathcal{N}_{\mathcal{G}}$ will have its Kraus operators which are not necessarily a linear combination of elements in $\mathbb{P}_n^X$. We call $\mathcal{N}_{X,\mathcal{G}}$ a CPTP map having a perfect bias (according to definition \textcolor{Blue}{4} of the main text) that approximates $\mathcal{N}_{\mathcal{G}}$. It allows to rewrite $\mathcal{N}_{\mathcal{G}}$ and $\mathcal{E}_{\mathcal{G}}$ as:
\begin{align}
&\mathcal{N}_{\mathcal{G}}=\mathcal{N}_{X,\mathcal{G}}+\Delta \mathcal{N}_{\mathcal{G}}\\
&\mathcal{E}_{\mathcal{G}}=\mathcal{E}_{X,\mathcal{G}}+\Delta \mathcal{E}_{\mathcal{G}},
\end{align}
with:
\begin{align}
\mathcal{E}_{X,\mathcal{G}} \equiv \mathcal{N}_{X,\mathcal{G}} \circ \mathcal{G}.
\end{align}

The smallest the diamond norm \cite{wilde2011classical} of $\Delta \mathcal{N}_{\mathcal{G}}$ will be, the better our approximation of a perfect bias will be. The property \ref{prop:diamond_perfect_bias} allows to quantify the maximum number of gates, $N$, that we can implement such that assuming a perfect bias does not make the measurement outcomes of the Hadamard test be further than an error budget $\epsilon$.
\begin{property}{Quality of the perfect-bias assumption}

    \label{prop:diamond_perfect_bias}
    We call $N$ the total number of gates used in the algorithm. We call $\mathcal{N}_{\mathcal{G}_i}$ the noise map describing the implementation of the $i$'th gate ($i \in [1,N]$) in the algorithm (the map describing the unitary implementation of this gate is labelled $\mathcal{G}_i$). We call $\mathcal{N}_{X,\mathcal{G}_i}$ a noise map approximating $\mathcal{N}_{\mathcal{G}_i}$ such that $\mathcal{N}_{X,\mathcal{G}_i}$ has a perfect bias according to definition \textcolor{Blue}{4} of the main text. We call $\rho_X$ the density matrix of the measured register if the noise model of gate $i$, for every $i$, was $\mathcal{N}_{X,\mathcal{G}_i}$. We call $\rho$ the final density matrix of the measured register if the noise model of gate $i$, for every $i$, was $\mathcal{N}_{\mathcal{G}_i}$. 
    If: 
    \begin{align}
    \max_i ||  \mathcal{N}_{X,\mathcal{G}_i}-\mathcal{N}_{\mathcal{G}_i}||_{\diamond} \leq \frac{\epsilon}{\sqrt{2} N},  
    \label{eq:max_diamond}
    \end{align}
    where $||.||_{\diamond}$ represents the diamond norm \cite{wilde2011classical}. Then, we can bound the error we make in the measurement outcome probability of the Hadamard test given our approximation of perfect bias. Specifically, for any single-qubit Pauli $P_1$,
    \begin{align}
        |Tr(\rho P_1)-Tr(\rho_X P_1)| \leq \epsilon
    \end{align}
\end{property}
\begin{proof} 
Calling $||.||_{H.S}$ the Hilbert-Schmidt norm (also called Frobenius norm \cite{coles2019strong}), from Cauchy-Schwarz inequality, we have: 
   \begin{align}
        |Tr(\rho P_1)-Tr(\rho_X P_1)| \leq ||\rho-\rho_X||_{H.S} ||P_1||_{H.S}.
    \end{align}
 Calling $||.||_1$ the trace norm, using the fact $||\rho-\rho_X||_{H.S} \leq ||\rho-\rho_X||_{1}$  \cite{coles2019strong}, and  $||P||_{H.S}=\sqrt{2}$, we also have:
    \begin{align}
        |Tr(\rho P_1)-Tr(\rho_X P_1)| \leq \sqrt{2} ||\rho-\rho_X||_{1}.
        \label{eq:diff_rho_rhoX}
    \end{align}
    For every $i \in [1,N]$, we define $\mathcal{E}_{\mathcal{G}_i} \equiv \mathcal{N}_{\mathcal{G}_i} \circ \mathcal{G}_i$ and $\mathcal{E}_{X,\mathcal{G}_i} \equiv \mathcal{N}_{X,\mathcal{G}_i} \circ \mathcal{G}_i$.
    Now, we assume that the noiseless algorithm implements the unitary $\mathcal{G}_{N} \circ ... \circ \mathcal{G}_1$. The noisy algorithm then implements: $\mathcal{E}_{\text{Algo}}=\mathcal{E}_{\mathcal{G}_N} \circ ... \circ \mathcal{E}_{\mathcal{G}_1}$. The noisy algorithm assuming the perfect bias assumption implements $\mathcal{E}_{X,\text{Algo}}=\mathcal{E}_{X,\mathcal{G}_N} \circ ... \circ \mathcal{E}_{X,\mathcal{G}_1}$. Calling $\rho_0$ the initial state for the whole algorithm (i.e. data, measured, and the potential parallelisation register of figure \ref{fig:parallelisation_register}), by definition of $\rho$ and $\rho_X$, we have:
    \begin{align}
        ||\rho-\rho_X||_{1}=&||Tr_{\neq \text{Measured}}[(\mathcal{E}_{\text{Algo}}-\mathcal{E}_{X,\text{Algo}})(\rho_0)]||_1 \notag \\
        &\leq ||(\mathcal{E}_{\text{Algo}}-\mathcal{E}_{X,\text{Algo}})(\rho_0)||_1 \leq ||\mathcal{E}_{\text{Algo}}-\mathcal{E}_{X,\text{Algo}}||_{\diamond}
        \label{eq:rho_to_diamond}
    \end{align}
    In \eqref{eq:rho_to_diamond}, $Tr_{\neq \text{Measured}}$ means that we trace out all degrees of freedom apart the measured register. We used the fact that trace distance is decreasing under partial trace, and the last inequality is implied by the definition of the diamond norm. Finally, using chaining properties of the diamond norm \cite{gilchrist2005distance}, we have:
    \begin{align}
        ||\mathcal{E}_{\text{Algo}}-\mathcal{E}_{X,\text{Algo}}||_{\diamond} \leq N \max_i ||\mathcal{E}_{\mathcal{G}_i}-\mathcal{E}_{X,\mathcal{G}_i}||_{\diamond}
        \label{eq:upper_bound_diff_E}
    \end{align}
    Combining \eqref{eq:diff_rho_rhoX}, \eqref{eq:rho_to_diamond}, \eqref{eq:upper_bound_diff_E}, and using the property of unitary invariance of the diamond norm (i.e. $||\mathcal{E}_{i}-\mathcal{E}_{X,\mathcal{G}_i}||_{\diamond}=||\mathcal{N}_{\mathcal{G}_i}-\mathcal{N}_{X,\mathcal{G}_i}||_{\diamond}$), we deduce:
    \begin{align}
        |Tr(\rho P)-Tr(\rho_X P)| \leq \sqrt{2}  N \max_i ||\mathcal{N}_{\mathcal{G}_i}-\mathcal{N}_{X,\mathcal{G}_i}||_{\diamond}
    \end{align}
    Hence, if $\max_i ||\mathcal{N}_{\mathcal{G}_i}-\mathcal{N}_{X,\mathcal{G}_i}||_{\diamond} \leq \epsilon/(\sqrt{2} N)$, then, $|Tr(\rho P)-Tr(\rho_X P)| \leq \epsilon$ which shows property \ref{prop:diamond_perfect_bias} holds.
    \end{proof}
\subsubsection{Benchmarking protocol}    
    \begin{restatable}[Benchmarking protocol]{theorem}{benchmarking}
\label{thm:benchmarking}
Consider a Hadamard test satisfying the constraints of Theorem \ref{app:thm:noise_resilience}. The circuit is implemented with $N$ unitary gates represented by the unitary quantum channels $\{\mathcal{G}_i\}_{i=1}^N$. We call $\mathcal{N}_{\mathcal{G}_i}$ the noise map associated to $\mathcal{G}_i$ (extracted from quantum tomography) and we assume state preparation and measurements to be noiseless (this is without loss of generality up to a redefinition of the noise maps of the quantum gates, see comments before Theorem \ref{thm:simulation_extended} in the main text).

Let $\mathcal{N}_{X,\mathcal{G}_i}$ be an approximation to $\mathcal{N}_{\mathcal{G}_i}$, such that $\mathcal{N}_{X,\mathcal{G}_i}$ has a noise model satisfying definition \textcolor{Blue}{4} of the main text.

Let $\rho$ be the density matrix of the measured register at the end of the algorithm if the noise map of each $\mathcal{G}_i$ was $\mathcal{N}_{\mathcal{G}_i}$. Let $\rho_X$ be the reduced density matrix of the measured register at the end of the algorithm if the noise map of each $\mathcal{G}_i$ was $\mathcal{N}_{X,\mathcal{G}_i}$. Let $\rho_{\text{exp}}$ be the reduced density matrix of the measured register that exactly predicts the experimental outcomes. More precisely, we mean that implementing the measurements used in the Hadamard test on $\rho_{\text{exp}}$ would exactly reproduces the measurement outcomes experimentally observed.

Assume that there exist $\epsilon>0$ such that
\begin{align}
    \max_{i} ||\mathcal{N}_{X,\mathcal{G}_i}-\mathcal{N}_{\mathcal{G}_i}|| \leq \frac{\epsilon}{\sqrt{2} N}
\end{align}
is satisfied. Then, for any single-qubit Pauli $P_1$:
\begin{align}
    |Tr(\rho P_1)-Tr(\rho_X P_1)| \leq \epsilon
\end{align}

\textbf{Principle of the benchmarking:}

The benchmarking protocol works as follow. $Tr(\rho_X P_1)$ corresponds to the outcome of the circuit with a noise model for each gate satisfying definition \textcolor{Blue}{4}. Hence, it can be classically estimated with Theorem \ref{thm:simulation_extended}. $Tr(\rho_{\text{exp}} P_1)$ is accessed experimentally. If $|Tr(\rho_{\text{exp}} P_1)-Tr(\rho_{X} P_1)|>\epsilon$, $\rho_{\text{exp}}$ and $\rho$ necessarily differ, indicating that the noise model predicted by individual gate tomography (the noise maps $\{\mathcal{N}_{\mathcal{G}_i}\}_{i=1}^N$) is not occuring experimentally. It thus indicates the presence of collective effects in the noise, possibly threatening the scalability of the biased-noise qubits.

Noteworthy, $\Delta \equiv |Tr(\rho_{\text{exp}} P_1)-Tr(\rho_X P_1)|-\epsilon$ can quantify how strong the noise violation is at the scale of the whole circuit (if $\Delta > 0$). It is a consequence of the fact $|Tr(\rho_{\text{exp}} P_1)-Tr(\rho P_1)| \geq \Delta$ (the larger $\Delta$, the larger the violation).
\end{restatable}

\begin{proof}
    The benchmarking protocol is a natural implication of property \ref{prop:diamond_perfect_bias}: if  
    \begin{align}
    \max_{i} ||\mathcal{N}_{X,\mathcal{G}_i}-\mathcal{N}_{\mathcal{G}_i}|| \leq \frac{\epsilon}{\sqrt{2} N},
\end{align}
then, from property \ref{prop:diamond_perfect_bias}, for any single-qubit Pauli $P_1$, we have: 
\begin{align}
    |Tr(\rho P_1)-Tr(\rho_X P_1)| \leq \epsilon
\end{align}
Hence, if $|Tr(\rho_{\text{exp}} P_1)-Tr(\rho_{X} P_1)|>\epsilon$, $\rho_{\text{exp}}$ and $\rho$ necessarily differ. Finally, using a simple triangular inequality, we obtain:
\begin{align}
    |Tr(\rho_{\text{exp}} P_1)-Tr(\rho_X P_1)| &\leq |Tr(\rho P_1)-Tr(\rho_X P_1)|\notag \\
    &+|Tr(\rho P_1)-Tr(\rho_{\text{exp}} P_1)| \notag \\
    & \leq \epsilon + |Tr(\rho P_1)-Tr(\rho_{\text{exp}} P_1)|
\end{align}
Hence $|Tr(\rho_{\text{exp}} P_1)-Tr(\rho P_1)| \geq \Delta$.
\end{proof}
\subsubsection{Estimating the size of implementable circuits based on literature}
    Our goal now is to estimate the maximum size of circuits for which we can implement our benchmarking protocol. For this purpose, we need to take concrete values for $\max_i ||\mathcal{N}_{\mathcal{G}_i}-\mathcal{N}_{X,\mathcal{G}_i}||_{\diamond}$ appearing in the Benchmarking protocol (theorem \ref{thm:benchmarking}) from the literature: it will allow us to deduce the total number of gates, state preparation we can do in the circuit, $N$, so that $\epsilon$ is sufficiently small to be useful. 
    
    We will represent the quantum channels through their $\chi$ matrix. Considering an $n$-qubit quantum channel $\mathcal{N}$, the $\chi$ matrix is the matrix of elements $\chi_{ij}$, where $\mathcal{N}(\rho)=\sum_{ij} \chi_{ij} P_i \rho P_j$, where $\{P_i\}$ is a family of $n$-qubit Pauli matrices forming a basis for the matrices living in an $n$-qubit Hilbert space. In \cite{xu2022engineering}, numerical simulations determined the $\chi$ matrix of the noise map of a cNOT. We will make our quantitative estimate based on this noise channel: our estimate should then only be taken as rough order-of-magnitudes estimates for the maximum circuit size allowed. Yet, the cNOT being typically noisier than a single-qubit gate, we think it is a well-motivated example to take for our numerics. One issue is that in \cite{xu2022engineering} (but also in other references, such as \cite{Guillaud2019Dec,Chamberland2022Feb}), the plots only provide the \textit{absolute} values for the real and imaginary parts of the off-diagonal elements of the $\chi$-matrix. For this reason, we perform the Pauli Twirling approximation (i.e. we neglect the off-diagonal terms of the $\chi$-matrix of the noise process), and we approximate the resulting channel by a perfectly biased Pauli channel. Our approximation could in principle under-estimate the actual value of the diamond distance, but this is the best we can do given the available information in literature. We can nonetheless notice that this approximation could in practice be well-justified in some cases, as the off-diagonal terms can be significantly lower than the dominant diagonal ones \cite{Chamberland2022Feb}. Hence, we consider:
\begin{align}
\mathcal{N}_{\text{cNOT}}(\rho)=\sum_{\substack{0 \leq i \leq 3\\ 0 \leq j \leq 3}} \chi^{\text{cNOT}}_{ijij} (\sigma_i \otimes \sigma_j) \rho (\sigma_i \otimes \sigma_j),
\label{eq:noise_channel_cNOT_from_literature}
\end{align}
where, from \cite{xu2022engineering} (we rounded the diagonal terms to the nearest order of magnitude from their color plot), we have:
\begin{align}
(\chi_{0000},\chi_{0303},\chi_{0202},\chi_{0101},\chi_{3030},\chi_{3333},\chi_{3232},\chi_{3131}, 
 \notag \\ \chi_{2020} 
,\chi_{2323},\chi_{2222},\chi_{2121},\chi_{1010},\chi_{1313},\chi_{1212},\chi_{1111})\notag\\
= \frac{1}{1.0012000066}(1, 10^{-9}, 10^{-10}, 10^{-4}, 10^{-9}, \notag \\ 10^{-9}, 10^{-10}, 10^{-10}, 10^{-9}, 10^{-9}, 10^{-10}, 10^{-10}, 10^{-3}, 10^{-9},\notag \\ 10^{-10}, 10^{-4}).
\label{eq:value_chi}
\end{align}
The term $1/1.0012000066$ is here to keep the map trace preserving (our rounding could make it lose this property). We draw the attention of the reader on the fact that in \eqref{eq:value_chi}, we took the values from \cite{xu2022engineering} but we permuted the roles of the Pauli matrices: $(I,X,Y,Z) \to (I,Z,-Y,X)$. This is because in \cite{xu2022engineering} they took the convention that the dominant noise mechanism was a phase-flip, while in our paper, it is a bit-flip. This "relabelling" physically correspond to a global change of basis through the application of a Hadamard gate to each qubit. We will approximate $\mathcal{N}_{\text{cNOT}}$ by the perfectly biased Pauli channel $\mathcal{N}_{X,\text{cNOT}}$:
\begin{align}
    \mathcal{N}_{X,\text{cNOT}}(\rho)=\sum_{\substack{0 \leq i \leq 3\\ 0 \leq j \leq 3}} \chi^{\text{cNOT}}_{X, ij ij} (\sigma_i \otimes \sigma_j) \rho (\sigma_i \otimes \sigma_j),
\end{align}
where:
\begin{align}
    (\chi^{\text{cNOT}}_{X, 0000},\chi^{\text{cNOT}}_{X, 0101},\chi^{\text{cNOT}}_{X, 1010},\chi^{\text{cNOT}}_{X, 1111})=\frac{1}{1.0012}(1,10^{-4},\notag \\ 10^{-3},10^{-4}),
\end{align}
all the other coefficients for $\chi^{\text{cNOT}}_{X, ijij}$ are equal to $0$. Using the analytical expression of the diamond distance between Pauli channels \cite{magesan2012characterizing}, we deduce $||\mathcal{N}_{\text{cNOT}}-\mathcal{N}_{X,\text{cNOT}}||_{\diamond} \approx 1.31 \times 10^{-8}$. Because we always have $|Tr(\rho P_1)-Tr(\rho_X P_1)| \leq 2$, a useful benchmark requires at least $\epsilon < 2$. Here, we consider a more conservative definition of useful benchmark by additionally asking that, for $Tr(\rho_X P_1)$ of order $\approx 1$, $Tr(\rho P_1)$ gives access to a reasonably close estimate to $Tr(\rho_X P_1)$, meaning that we consider $\epsilon \ll 1$. Loosely speaking, it means we ask our noise approximation to closely represent the measurement outcomes we would have obtained with the expected noise model (i.e. the one deduced from individual gate tomography). Considering for instance $\epsilon=1/50$, property \ref{prop:diamond_perfect_bias} provides us:
\begin{align}
    N \leq 1.07 \times 10^{6}.
\end{align}
Note that it implicitly assumes that $||\mathcal{N}_{\text{cNOT}}-\mathcal{N}_{X,\text{cNOT}}||_{\diamond}$ gives a fair estimate to the left handside of \eqref{eq:max_diamond} (we believe it to be fair as a cNOT is typically noisier than single-qubit gates). Hence, our benchmark is applicable for circuits composed of at most $1.07 \times 10^{6}$ gates. This is $3$ to $4$ orders of magnitude larger than what hardware and circuits not exploiting a noise bias can implement \cite{pelofske2022quantum,Sung2021Jun}. Our benchmark is then in practice applicable for near and longer-term circuits.
\bibliography{supp}